\documentclass[11pt]{article} 
\makeatletter\renewcommand{\section}{\@startsection
	{section}{1}{\z@}{-3.5ex plus -1ex minus
		-.2ex}{2.3ex plus .2ex}{\bf }}
\makeatletter\renewcommand{\subsection}{\@startsection{subsection}{2}{\z@}{-3.25ex
		plus -1ex minus
		-.2ex}{1.5ex plus .2ex}{\it }}
\makeatletter\renewcommand{\subsubsection}{\@startsection{subsubsection}{3}{-2.45ex}{-3.25ex
		plus -1ex minus -.2ex}{1.5ex plus .2ex}{\it }}
\renewcommand{\thesection}{\arabic{section}}
\renewcommand{\thesubsection}{\arabic{section}.\arabic{subsection}.}

\renewcommand{\theequation}{\thesection.\arabic{equation}}
\makeatletter \@addtoreset{equation}{section}
\usepackage[table,xcdraw]{xcolor}
\renewenvironment{thebibliography}[1]
{\baselineskip=16pt plus 2pt minus 1pt
	\section*{\large\refname
		\@mkboth{\MakeUppercase\refname}{\MakeUppercase\refname}}%
	\list{\@biblabel{\@arabic\c@enumiv}}%
	{\settowidth\labelwidth{\@biblabel{#1}}%
		\leftmargin\labelwidth
		\advance\leftmargin\labelsep
		\@openbib@code
		\usecounter{enumiv}%
		\let\p@enumiv\@empty
		\renewcommand\theenumiv{\@arabic\c@enumiv}}%
	\sloppy
	\clubpenalty4000
	\@clubpenalty \clubpenalty
	\widowpenalty4000%
	\sfcode`\.\@m}

\let\fn\footnote
\renewcommand{\footnote}[1]{\linespread{1.1}\fn{#1}\linespread{1.29}}

\usepackage{multirow}
\usepackage{epsfig} 
\usepackage{verbatim}
\usepackage{graphicx}
\usepackage{tikz}
\usetikzlibrary{trees,er,snakes,shapes,mindmap}
\usepackage{hyperref}
\usepackage{easyfig}
\usepackage{bbm}
\usepackage{cite}
\usepackage{slashed}
\usepackage{physics}
\usepackage{amsmath}
\usepackage{amsfonts}
\usepackage{amssymb}
\usepackage{braket}
\usepackage{mathrsfs}
\usepackage{bm}
\usepackage[section]{placeins}
\usepackage{placeins}
\usepackage{mathtools}
\usepackage[utf8x]{inputenc}
\usepackage{booktabs}
\usepackage{caption}

\newcommand{\appendices}{\section*{Appendices}\setcounter{section}{0} \setcounter{equation}{0}
	\renewcommand{\thesection}{\Alph{section}.}
	\renewcommand{\thesubsection}{\Alph{section}.\arabic{subsection}.}
	\renewcommand{\theequation}{\thesection\arabic{equation}}}

\def\tyng(#1){\hbox{\tiny$\yng(#1)$}}

\newcommand{\del}{\partial}

\let\Oldsection\section
\renewcommand{\section}{\FloatBarrier\Oldsection}

\let\Oldsubsection\subsection
\renewcommand{\subsection}{\FloatBarrier\Oldsubsection}

\let\Oldsubsubsection\subsubsection
\renewcommand{\subsubsection}{\FloatBarrier\Oldsubsubsection}

\allowdisplaybreaks
\newcommand{\be}{\begin{equation}}
\newcommand{\ee}{\end{equation}}
\newcommand{\beq}{\begin{equation}}
\newcommand{\eeq}{\end{equation}}
\newcommand{\bea}{\begin{array}}
	\newcommand{\ea}{\end{array}}
\newcommand{\beqa}{\begin{eqnarray}}
\newcommand{\eeqa}{\end{eqnarray}}
\newcommand{\beqar}{\begin{eqnarray}}
\newcommand{\eeqar}{\end{eqnarray}}
\newcommand{\nn}{\nonumber}

\def\half{\textstyle{1\over 2}}
\DeclareMathOperator{\Ree}{\mathfrak{Re}}

\begin{document}
\fontfamily{bch}\fontsize{11pt}{15pt}\selectfont
	\begin{titlepage}
		\begin{flushright}
		\end{flushright}
		
		\vskip 2 em
		
		\begin{center}
{\Large \bf Magnetic Field and Curvature Effects on Pair Production}\\
~\\
{\Large \bf II: Vectors and Implications for Chromodynamics}
			
			\vskip 1.5cm
			
			\centerline{$ \text{\large{\bf{D. Karabali}}}^{a} $, $ \text{\large{\bf{S. K\"{u}rk\c{c}\"{u}o\v{g}lu}}}^{b,c} $, $ \text{\large{\bf{V.P.Nair}}}^{c} $}
			
			\vskip 0.5cm
			\centerline{\sl $^a$ Department of Physics and Astronomy}
			\centerline{\sl Lehman College of the CUNY, Bronx, NY, 10468,USA}
			\vskip 1em
			\centerline{\sl $^b$ Middle East Technical University, Department of Physics,}
			\centerline{\sl Dumlupinar Boulevard, 06800, Ankara, Turkey}
			\vskip 1em
			\centerline{\sl $^c$ Physics Department City College of the CUNY   }
			\centerline{\sl New York, NY 10031 USA}
			\vskip 1em
 \vskip .26cm
\begin{tabular}{r l}
E-mail:&\!\!\!{\fontfamily{cmtt}\fontsize{11pt}{15pt}\selectfont dimitra.karabali@lehman.cuny.edu}\\
&\!\!\!{\fontfamily{cmtt}\fontsize{11pt}{15pt}\selectfont kseckin@metu.edu.tr}\\
&\!\!\!{\fontfamily{cmtt}\fontsize{11pt}{15pt}\selectfont vpnair@ccny.cuny.edu}
\end{tabular}
			
		\end{center}
		
		\vskip 4 em
		
		\begin{quote}
			\begin{center}
				{\bf Abstract}
			\end{center}
			
			\vskip 1em
			
We calculate the
pair production rates for spin-$1$ or vector particles
on spaces of the form $M \times {\mathbb R}^{1,1}$
with $M$ corresponding to ${\mathbb R}^2$ (flat), $S^2$
(positive curvature) and $H^2$ (negative curvature),
with and without a background (chromo)magnetic field on $M$.
Beyond highlighting the effects of curvature and background
magnetic field, this is particularly interesting since vector particles are
known to suffer from the Nielsen-Olesen instability, which can dramatically
increase pair production rates.
The form of this instability for $S^2$ and $H^2$ is obtained.
We also give a brief discussion of how our results relate to
ideas about confinement in nonabelian theories.

			\vskip 1em
			
			\vskip 5pt
			
		\end{quote}
		
	\end{titlepage}
\setcounter{footnote}{0}
\pagestyle{plain} \setcounter{page}{2}
	
\newpage

\section{Introduction}

In a previous paper we have analyzed the Schwinger pair production process on spacetimes
with nonzero curvature and with a background magnetic field, in addition to the uniform
electric field \cite{KKN1}.
The general motivation for this was to elucidate the impact of spin-curvature and 
spin-magnetic field
couplings on threshold effects and hence on pair production rates.
We considered cases of the spacetime manifolds which would provide
explicit solvable examples. The analysis in \cite{KKN1} was for spin-zero and 
spin-$\half$
particles. Specifically, in addition to flat Minkowski space
${\mathbb R}^{3,1}$, we considered the manifolds
$M \times {\mathbb R}^{1,1}$ with $M = \, S^2, \, H^2, \, T^2$,
$S^2$ being the two-sphere, $H^2$ the 
two-dimensional hyperbolic plane and $T^2$ the flat two-torus.
In the present paper, we will consider similar analyses for the
spin-$1$ particles. The Schwinger process has a long history. For the proper placement of
our work and for some of the 
relevant literature, we refer the reader to references cited in
\cite{KKN1}.

There are many issues which make the discussion of spin-$1$ particles 
more involved compared to the scalar and spinor cases
and there are some interesting and new facets as well.
There are well known no-go theorems which point to
difficulties in constructing a field theory of charged spin-$1$ 
particles \cite{WW}. The only consistent formulation is to
consider the spin-$1$ charged fields as part of a nonabelian gauge field,
minimally an $SU(2)$ gauge theory. One of the nonabelian fields, say
the $A^3_\mu$ field for the $SU(2)$ example, can be considered as
the electromagnetic field and the other components as the charged spin-$1$
 fields.
 In this case, the Yang-Mills action automatically incorporates the correct
 spin-magnetic field and spin-curvature couplings, the gyromagnetic ratio being
 $2$, as it is for Dirac particles. We also have to ensure that only the correct
 number of physical polarizations are effective in the pair production process.
 This will require the elimination of the unphysical polarizations
 via gauge fixing
 or via suitable constraints on the quantum states.
 The simplest way will be to use the BRST formalism, 
 which is what we do in what follows.
 
Since the spin-$1$ fields are part of a Yang-Mills multiplet, our analysis has the potential
to generate results relevant to QCD; this is another motivation for the present work.
For Yang-Mills fields, it has been known for a long time that
the vacuum tends to generate chromomagnetic fields \cite{{savv}, {F^2}}
and at the same time that
there is an instability for such fields \cite{{NO}, {mass}}.
The decay of chromoelectric fields via the Schwinger effect has also been considered
in \cite{NY}, where it was argued that, for a nonabelian
plasma, there is an
instability which does not 
allow for a net nonzero color charge along with
field configurations which are coherent over a length scale given by 
the chemical potential. Therefore the situation with both chromoelectric and chromomagnetic 
fields is clearly an interesting case.

This paper is organized as follows. In section 2, we set up the basic framework and the calculation of
the rate for pair production in flat Minkowski space ${\mathbb R}^{3,1}$.
 By taking the limit of this result for
zero electric field we are also able to recover the Nielsen-Olesen result \cite{NO}.
The following two sections are devoted to similar calculations for 
$S^2 \times {\mathbb R}^{1,1}$ and $H^2 \times {\mathbb R}^{1,1}$.
In section 5, we give a brief summary and also
consider our results in the context of chromodynamics, 
commenting on the difficulties of maintaining stable chromoelectric configurations.
Calculation of the energy spectrum for the charged vector particles and the corresponding density of states on the hyperbolic plane require the use of representation theory of $SL(2, {\mathbb R})$. 
Essential points regarding this as well as a semiclassical estimate of the density of states on $S^n$ and $H^n$, using the Hamilton-Jacobi theory  
are provided in the appendices A and B, respectively.

\section{Pair production of vector particles in flat space}

We launch our discussion by considering vector particles in flat space. For this we need the action for a charged vector field coupled to background magnetic fields, including the magnetic moment or Zeeman coupling term. We have already seen, for the case of spin-$\half$ particles, that the Zeeman coupling has a crucial role in enhancing pair production via the zero modes \cite{KKN1}. The only consistent theory for charged vector fields must treat them as part of a nonabelian multiplet. Thus we start with an $SU(2)$ gauge theory with the dynamics given by the Yang-Mills action. Some of the arguments we develop can be applied to QCD as well, so the Yang-Mills theory is indeed the appropriate starting point. The Euclidean Yang-Mills action is given by
\beq
S = {1\over 4}\int d^4x \, F^a_{\mu\nu} F^a_{\mu\nu} = - {1\over 2}
\int d^4x\, \mbox{Tr} (F_{\mu\nu} F_{\mu\nu} )
\label{rev18}
\eeq
where, as usual, $F_{\mu\nu} = (-i t^a) F^a_{\mu\nu} = \del_\mu A_\nu - \del_\nu A_\mu + [A_\mu , A_\nu ]$ and $A_\mu = (-it^a) A^a_\mu$ is the gauge potential. $\{ t^a\}, a = 1, 2, 3$, are hermitian matrices forming a basis for the Lie algebra
of $SU(2)$; thus $t^a$ obey the usual commutator algebra with the structure constants
$f^{abc}$, i.e., 
$[t^a , t^b] = i f^{abc} t^c$, and we take them to be normalized as $\Tr ( t^a t^b ) = \half \delta^{ab}$.

We introduce a background for the $SU(2)$ gauge potential by $A^a_\mu \rightarrow A^a_\mu + W^a_\mu$ where $A^a_\mu$ is now the background field and $W^a_\mu$ denote the fluctuations around the background. The problem of gauge-fixing and reduction to the physical polarizations can be dealt with using the BRST formalism. The BRST transformations are given by
\beqar
Q (A^a_\mu + W^a_\mu) &=& \del_\mu c^a + [A_\mu + W_\mu, c]^a
= (D_\mu c)^a + [W_\mu, c]^a\nonumber\\
Q\, c^a &=& - {1\over 2} f^{abc} \, c^b c^c\nonumber\\
Q \, {\bar c}^a &=& b^a, \hskip .2in Q\, b^a = 0
\label{rev19}
\eeqar
Here $D_\mu$ denotes the covariant derivative with the background field
$A_\mu$ as the connection and $b^a$ is the Nakanishi-Lautrup field.
Since this is a fixed background, $Q A^a_\mu = 0$, and the first of the equations 
in (\ref{rev19}) is to be interpreted as
$Q\,W^a_\mu  = (D_\mu c)^a + [W_\mu, c]^a$.
We take the gauge-fixed action to be
\beqar
S&=& S_{\rm YM} (A+W) + Q \left[\int {\bar c}^a \left( (D_\mu W^\mu)^a -{\half}  b^a \right)
\right]\nonumber\\
&=&S_{\rm YM} (A+W)  + \int d^4x\,\left[ (D_\mu {\bar c})^a \, (D^\mu c + [W^\mu, c])^a 
+ {1\over 2}  \, (D\cdot W )^a (D\cdot W)^a\right]
\label{rev20}
\eeqar
We have done some partial integrations and also, in the second line,
eliminated $b^a$ by its equation of motion,
which is equivalent to integrating it out in the functional integral.
The Yang-Mills part of the action simplifies to
\beqar
S_{\rm YM} (A + W ) &=& {1\over 4}\int d^4x\, f_{\mu\nu}^a f^{a\mu\nu} +  {\tilde S} \nonumber\\
{\tilde S}&=& \int d^4x\, \Bigl[
{1\over 2} (D_\mu W_\nu )^a  (D^\mu W^\nu )^a - {1\over 2} (D\cdot W)^a (D\cdot W)^a
+ f^{abc} \, f^a_{\mu\nu} W^b_\mu W^c_\nu \nonumber\\
&& \hskip .5in + {\rm cubic ~and ~quartic ~terms~ in~} W
\Bigr]\label{rev21}
\eeqar
Here $f_{\mu\nu}$ is the field strength tensor for the background field $A_\mu$.
We have omitted the term linear in $W$ as it vanishes for backgrounds which
obey the classical equations of motion.
The term $f^{abc} \, f^a_{\mu\nu} W^b_\mu W^c_\nu$ is the Zeeman coupling corresponding to
a gyromagnetic ratio of $2$.
Combining (\ref{rev20}) and (\ref{rev21}), we find
\beqar
S&=&  \int d^4x\, \Bigl[
{1\over 2} (D_\mu W_\nu )^a  (D^\mu W^\nu )^a 
+ f^{abc} \, f^a_{\mu\nu} W^b_\mu W^c_\nu
+(D_\mu {\bar c})^a \, (D^\mu c)^a \nonumber\\
&& \hskip .5in +\, {\rm cubic ~and ~quartic ~terms~ in~} W, {\bar c}, c
\Bigr]\label{rev22}
\eeqar
We take the background field to be along the $t^3$ direction in the $SU(2)$ algebra, so
that
\beq
f^3_{12} = - f^3_{21} = B_1, \hskip .2in
f^3_{34} = - f^3_{43} = B_2
\label{rev23}
\eeq
Further, we take combinations of $W^a_\mu$ and the ghosts to define the fields
\beqar
W_+^+ = {1\over 2} \left[ W_1^1 - W_2^2 + i (W_2^1 + W_1^2) \right],~~&&~~
W_+^- = {1\over 2 } \left[ W_1^1 + W_2^2 + i (W_2^1 - W_1^2) \right]\nonumber\\
w_+^+ = {1\over 2} \left[ W_3^1 - W_4^2 + i (W_4^1 + W_3^2) \right],~~&&~~
w_+^- = {1\over 2 } \left[ W_3^1 + W_4^2 + i (W_4^1 - W_3^2) \right]\nonumber\\
c^\pm = {1\over \sqrt{2}} (c^1 \pm i c^2), ~~&&~~
{\bar c}^\pm = {1\over \sqrt{2}} ({\bar c}^1 \pm i {\bar c}^2)
\label{rev24}
\eeqar
The fields $W^3_\mu$, $c^3$, ${\bar c}^3$ are uncharged with respect to the background field.
Further $W_-^- = W_+^{+*}$, $W_-^+ = W_+^{-*}$, etc.
The quadratic terms in the action (\ref{rev22}) become
\beqar
S &=& \int d^4x \Bigl[ (D_\mu W_+)^{+*} (D^\mu W_+)^+ + (D_\mu W_-)^{+*} (D^\mu W_-)^+ 
- 2 B_1 ( W_+^{+*} W_+^+ - W_-^{+*} W_-^+ )\nonumber\\
&& \hskip .5in + (D_\mu w_+)^{+*} (D^\mu w_+)^+ + (D_\mu w_-)^{+*} (D^\mu w_-)^+ 
- 2 B_2 ( w_+^{+*} w_+^+ - w_-^{+*} w_-^+ )\nonumber\\
&&\hskip .5in +
(D_\mu {\bar c})^- (D^\mu c)^+ + (D_\mu {\bar c})^+ (D^\mu c)^-
+ {1\over 2} \del_\mu W_\nu^3 \,\del^\mu W^{3\nu}
+ \del_\mu {\bar c}^3 \del^\mu c^3\Bigr]
\label{rev25}
\eeqar

\begin{table}[!t]
	\caption{Eigenvalues for the charged components of the vector fields and ghosts}  
	\label{Flat-eigen}
	\small 
	\centering 
	\begin{tabular}{ccl} 
		\toprule[\heavyrulewidth]\toprule[\heavyrulewidth]
		\textbf{Fields} & \textbf{Eigenvalues} & \textbf{} \\ 
		\midrule
$c^\pm$&{$(2 n_1+1) B_1+$} & {$ \hspace{-1em}(2 n_2 +1) B_2$} \\
\rule{0pt}{4ex} 
$w_-^+$& \multirow{2}{*}{$(2 n_1+1) B_1$ +} & $ \hspace{-1em} (2 n_2 +3) B_2$ \\
$w_+^+$& & $  \hspace{-1em} (2 n_2 -1) B_2$\\
\rule{0pt}{4ex} 
$W_-^+$&$(2 n_1+3) B_1$  & \multirow{2}{*}{$ \hspace{-1em} + (2 n_2 +1) B_2$}\\
$W_+^+$&$ (2 n_1-1) B_1$  & \\
\bottomrule[\heavyrulewidth] 
\end{tabular}
\end{table}
The eigenvalues of the kinetic operator (which we shall often refer to as the energy eigenvalues
even though we are discussing the Euclidean action)
for $W^3_\mu$ and $c^3$, ${\bar c}^3$ are independent of the magnetic 
fields and lead to an infinite constant in the effective action 
which is removed by renormalization. The eigenvalues for the charged fields are shown in
Table \ref{Flat-eigen}. The density of states is $(B_1 B_2/ 4 \pi^2)$, for all cases.
Using these values,
we find the effective action $\Gamma = \Gamma_1 + \Gamma_2$ with
\beqar
- \Gamma_1 &=&  {1\over 8\pi^2} \int d^4x {d s \over s}
 \left( {B_2 \over \sinh s B_2}\right) \left[
 B_1 \sum_{n_1} \left( e^{- s [m^2 + ( 2n_1 -1) B_1] } + e^{- s ( 2n_1 +3 ) B_1} \right)\right]
 \nonumber\\
 - \Gamma_2 &=&  {1\over 4\pi^2} \int d^4x {d s \over s}
 \left( {B_2  \over \sinh sB_2}\right)(\cosh 2sB_2 \, -1)  \left[
 B_1 \sum_{n_1}  e^{- s ( 2n_1 +1) B_1} \right]
 \label{rev26}
 \eeqar
 $\Gamma_1$ is the contribution from $W_\pm^+$. 
 Rather than a zero mode as in the case of spin-$\half$, we now have a negative
 mode 
 $(2n_1 + 1)B_1 - 2 B_1 = - B_1$ (for $n_1 = 0$),
 due to the Zeeman term. This instability is what was noticed long ago
 by Nielsen and Olesen \cite{NO} and has led to arguments in favor of
 the QCD vacuum spontaneously generating chromomagnetic fields with a consequent
 instability which is then eliminated by mass generation.
 We will discuss the physics of this later.
 For now, we notice that the negative mode can lead to
 a divergence, so, we have introduced a mass term as an ad hoc infrared cutoff.

The rate for pair production or decay of the field is obtained by continuing
$\Gamma$ to Minkowski signature
by using
$x_4 \rightarrow i x_0$, $B_2 \rightarrow -i E $. The continuation of
$-\Gamma$ should then be identified as
$i S_{\rm eff}$,
with the vacuum-to-vacuum amplitude given as
$\braket{0|0} = e^{i S_{\rm eff}}$. We are thus interested in the real part of $i S_{\rm eff}$.
The factor $ (\sinh s B_2)^{-1}$  in (\ref{rev26}), upon continuation, becomes
$(- i \sin Es)^{-1}$ and can potentially produce poles at 
$s E = n  \pi$, $n = 1, 2, \cdots$.
(There is no singularity at $n =0$,  or $s =0$, since the
integration over $s$ starts at $s = \epsilon$, with $\epsilon$ being an ultraviolet cut-off.
To put another way,
the $s =0$ singularity is subtracted out
via renormalization before taking $\epsilon \rightarrow 0$.)
The imaginary part of $S_{\rm eff}$ (i.e., the real part of $i S_{\rm eff}$)
arises from going around the poles in doing the $s$-integration.
Near $s E = n \pi$, we write
$s = (n \pi /E) + z$, $\sin s E = \sin (n \pi + E z) \simeq (-1)^n E z$
and
calculate the contribution from integration over a small semicircle around these points
to obtain
$\Ree (i S_{\rm eff})$ \cite{KKN1}.

The second term in the effective action, namely, $\Gamma_2$, is the contribution from $w$'s and the ghosts.
In going over to Minkowski space, we note that
 since
$\cosh 2 s B_2 - 1= 2 \sinh^2 sB_2$, the continuation of
$\Gamma_2$ to Minkowski signature does not have any
poles and so it
does not contribute to the pair production. This is a reflection of the
ghosts cancelling out the effects of two polarizations of the vector particle, reducing the
physics to that of the two physical polarizations.

From $\Gamma_1$ , which is due to $W_\pm^+$, we get 
\beqar
\Ree (i S_{\rm eff}) &=&  \int d^4x\, {E \over 8 \pi^2} 
\sum_{n =1}^\infty {(-1)^n \over n} \,
 B_1 \sum_{n_1} \left( e^{- s [ m^2+ ( 2n_1 -1) B_1]} + e^{- s ( 2n_1 +3 ) B_1} \right)_{s= n\pi/E}
 \nonumber\\
 &=& - \int d^4x\, {E^2 \over 96 \pi} \, f_1(B_1/E)\label{rev27}\\
 f_1(x)&=& {12 \,x \over \pi} \sum_{n =1}^\infty {(-1)^{n+1} \over n} 
 \sum_{n_1} \left( e^{- [ m^2 (n \pi /E) + (2 n_1 -1) n \pi x] } +
 e^{- (2 n_1 +3 ) n \pi x }  \right)\nonumber\\
 &=&{12 \,x \over \pi} \Bigl[ \log \bigl( 1+ e^{\pi x}\bigr) + \log \bigl( 1+ e^{-\pi x}\bigr) 
 + 2\, \sum_{n_1=1}^\infty \log \bigl(1 + e^{ - (2 n_1 +1 ) \pi x}\bigr)\Bigr]
 \label{rev28}
 \eeqar
 The summation over $n$ in $f_1 (x)$ will converge if $m^2 > B_1$.
 We have used this to calculate the sum and then set $m^2 =0$.
 This may be viewed as defining the sum in a region where it converges 
 and then defining the result 
 in the larger domain by continuation. 
 For zero magnetic field, we have $f_1 = 1$, so enhancement effects can be
 identified by considering values of $f_1 (x)$. 
A graph of this function is shown in Fig.\,{\ref{Flat-vec1}.
We see clearly  that there is enhancement of pair production due to 
the magnetic field, namely, $f_1(x) \geq 1$.
\vskip .2in
\begin{figure}[!htb]
\begin{center}
\begin{tikzpicture}[scale = 1.4,domain=0:7]
\pgftext{
\scalebox{.4}{\includegraphics{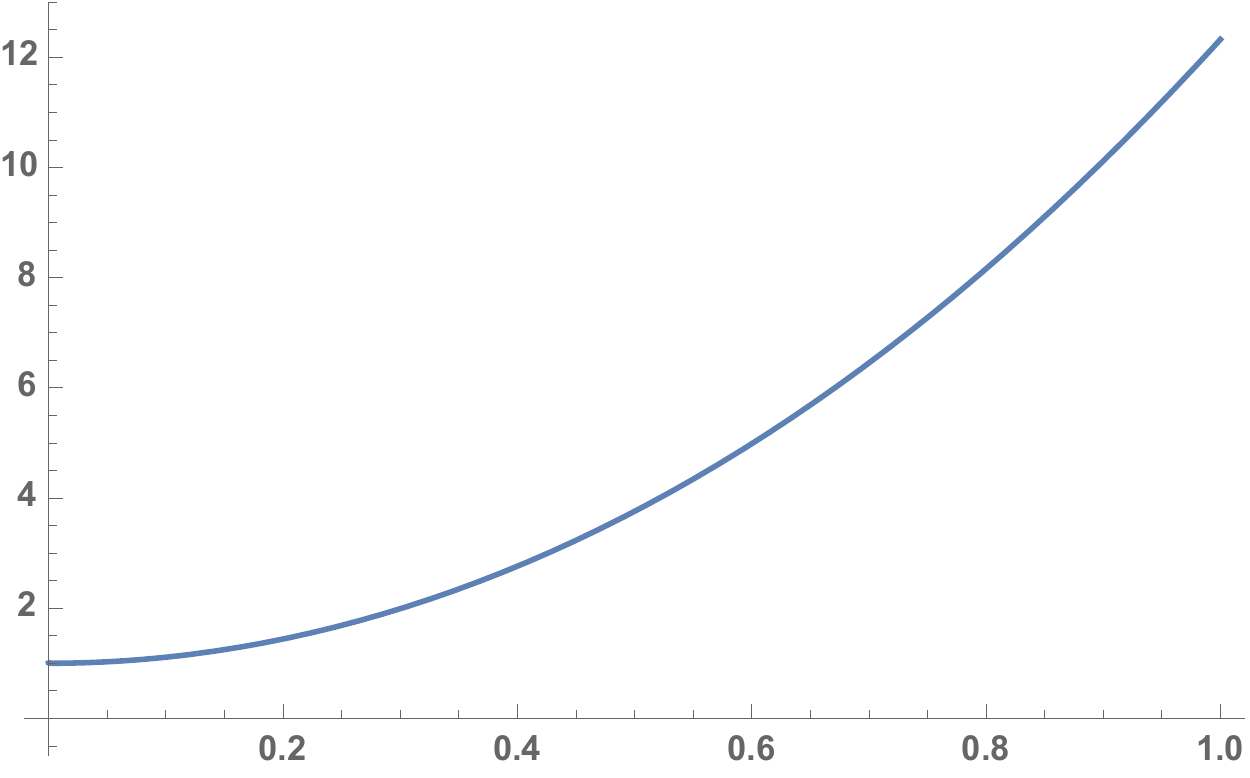}}
}
\draw(-3.0,.5)node{$f_1(x)$};
\draw(1,-1.7)node{$x \rightarrow$};
\end{tikzpicture}

\caption{The graph of $f_1(x)$ showing enhancement due to the background magnetic field}
\label{Flat-vec1}
\end{center}
\end{figure}

It is also interesting to consider the large $x$ limit of $f_1 (x) \approx 12\, x^2$
from (\ref{rev28}).
In this  limit, which also corresponds to
$E \rightarrow 0$, we find
\beq
\Ree (i S_{\rm eff}) = - \int d^4x\,{B_1^2 \over 8 \pi}
\label{rev28a}
\eeq
This is exactly the result for the magnetic instability given in
\cite{NO}. Thus equations (\ref{rev27}), (\ref{rev28}) do incorporate the Nielsen-Olesen instability.
		
\section{Pair Creation of Vector Particles on $S^2 \times {\mathbb R}^{1,1}$}

The case of vector particles on $S^2 \times {\mathbb R}^{2}$
can be analyzed using what we did in flat space, with only a couple
of changes. We will consider uniform magnetic fields on 
$S^2$ and ${\mathbb R}^{2}$, which we take to be of the form
$ B_1 t^3$ and $B_2 t^3$, respectively.
On the sphere there is a quantization condition for the field $B_1$; it can be written as
$B_1 = \frac{N}{2 a^2}$, 
with $N \in {\mathbb Z}$, $a$ being the radius of curvature of the sphere.
The wave functions for the $S^2$ part will be given by
representation matrices for $SU(2)$ of the form
$\bra{j, m} {\hat g} \ket{j, m'}$, for ${\hat g} \in SU(2)$, where
we regard $S^2$ as $ SU(2)/U(1)$.
(The $SU(2)$ here is used to define the manifold $S^2$ and it
is not related to the gauge group which we have chosen
to be $SU(2)$ as well.)
The Laplace operator now takes the form
of $-D_3^2 - D_4^2 + (R_1^2 + R_2^2 )/a^2$, where $R_a$ are the
right translation operators acting on ${\hat g}$.
$i R_1/a$, $i R_2/a$ define the covariant derivatives on the sphere,
the commutator of which should be
the magnetic field
(multiplied by the charge matrix) plus the curvature (multiplied by the spin).
Since
\beq
[D_1 , D_2] = - {[R_1, R_2]\over a^2} = -i {R_3 \over a^2}
\label{Sp-vec0}
\eeq
we see that the eigenvalue of $R_3$ must be 
chosen to represent the sum of the magnetic field and the appropriate
spin-curvature term.
The eigenvalues
of $R_1^2 + R_2^2 $ can be evaluated by
writing it as $(R_1^2 + R_2^2 + R_3^2) - R_3^2 = j(j+1) - R_3^2$, with $R_3$ set to the appropriate values
for 
the various components keeping in mind that
$w_\pm$ will be scalars on $S^2$, while $W_\pm$ are vectors.
This means that we must have the $R_3$ and $j$ values as given in
Table\,\ref{R3}, where $q$ is an integer, $q \geq 0$.
\begin{table}[!htb]
	\begin{center}
		\caption{$R_3$ and $j$ values for the charged fields}
\label{R3}
		\begin{tabular}{c c  c l}
			\toprule[\heavyrulewidth]\toprule[\heavyrulewidth]
			\textbf{Fields} & {$\mathbf R_3$} &  $\mathbf j$ &~ \\ 
			\midrule 	
			$c^\pm$, $w^+_\pm$& $- {N\over 2}$& $ q + {N \over 2}$&\\
			\rule{0pt}{4ex} 
			$W_-^+$&$-1 - {N \over 2} $& ~~~$q + 1+  {N\over 2}$~~~&\\
			\rule{0pt}{4ex} 
			$W_+^+$& $1 - {N\over 2}$&$q - 1+ {N\over 2} $ &{\rm if}~~$N\geq 2$\\
			&${1\over 2}$&$q + {1\over 2}$ &{\rm if}~~$N =1$\\
			&$1$& $q+1$&{\rm if}~~$N =0$\\

\bottomrule[\heavyrulewidth] 
		\end{tabular}
	\end{center}
\end{table} 
The degeneracy factor for the Landau levels on the sphere will
be $2 j +1$ as usual.
(For general background on such calculations, see
\cite{KN}.)

In the action, there is also an additional Zeeman-like term due to the coupling of
spin to curvature. We can simplify the action as follows.
With the background fields,
we have 
\beq
F_{\mu\nu} = f_{\mu\nu} + (D_\mu W_\nu - D_\nu W_\mu ) + [W_\mu, W_\nu ]
\label{Sp-vec1}
\eeq
In using this in (\ref{rev18}), the quadratic terms are given by
\beqar
S(A+ W) &=& \int d^4x \left[
-\Tr (D_\mu W_\nu \, D_\mu W_\nu) + \Tr ( D_\mu W_\nu \, D_\nu W_\mu )
- \Tr (f_{\mu\nu} [W_\mu, W_\nu]) \right] + \cdots\nonumber\\
&=&\int d^4x \left[ -\Tr ( W_\nu (-D^2 )W_\nu ) - \Tr (W_\nu [D_\mu, D_\nu]W_\mu)
+ \Tr (D\cdot W )^2\right.\nonumber\\
&&\hskip .5in \left. - \Tr (f_{\mu\nu} [W_\mu, W_\nu]) \right] + \cdots
\label{Sp-vec2}
\eeqar
In the case of flat space, the commutator of the two covariant derivatives acting
on $W_\mu$ gave $[f_{\mu\nu}, W_\mu]$ which added on to the last term and
gave the Zeeman term $- 2 \Tr f_{\mu\nu} [W_\mu, W_\nu]$.
As mentioned above, the commutator also gives a term proportional to
the Riemann tensor. For $S^2 \times {\mathbb R}^2$, this applies to the case
of $\mu, \nu = 1,2$. Using $D_b = i R_b /a$, $b= 1,2$,
this term can be evaluated as
\beqar
- \Tr (W_\nu [D_\mu, D_\nu]W_\mu)
&=& {1\over a^2}  \Tr (W_b [R_a, R_b]W_a)
=  {i\over a^2}  \Tr ( - W_1 R_3W_2 + W_2 R_3 W_1)\nonumber\\
&=&{1\over a^2} \Tr ( W_+ R_3 W_-  -  W_- R_3 W_+ )\nonumber\\
&=&  {1\over 2 a^2} \bigl( W^c_- R_3 W^c_+ - W^c_+ R_3 W^c_-)
= {1\over a^2} \, W^c_- W^c_+
\label{Sp-vec3}
\eeqar
In this equation, in evaluating the action of $R_3$, 
we have only indicated the gravitational or curvature part since
 the gauge field part involving $f_{\mu\nu}$ was already included
 (as part of the Zeeman term).
 It is now easy to write down the spectrum of various fields.
 Ignoring the uncharged fields $W_\mu^3, {\bar c}^3, c^3$, which do not contribute to
 the pair production, the fields and the corresponding eigenvalues and densities of states $\rho$ are as 
 given in Table \ref{sphere-eigen}. We have not indicated the conjugate fields.

\begin{table}[!htb]
	\begin{center}
		\caption{Eigenvalues and degeneracies for the charged components of the vector fields
			and ghosts}
\label{sphere-eigen}
		\begin{tabular}{c c l l}
			\toprule[\heavyrulewidth]\toprule[\heavyrulewidth]
			\textbf{Fields} &&\hspace{-1em}\textbf{Eigenvalues}  & $\bm{ 8 \pi^2 a^2 \, \rho}$  \\ 
			\midrule 
$c^\pm$&{$(2 n_2 +1 )B_2 $} &\hspace{-1em}{$~+~\bigl( q (q+1) + N (q + {\half})\bigr)/a^2$} &{$B_2 {( N + 2 q +1)} $}\\	
\rule{0pt}{4ex}
			$w_-^+$& $(2 n_2 +3) B_2$ & \multirow{2}{*}{$ \hspace{-1em}~+~ \bigl( q (q+1) + N (q + {\half})\bigr)/a^2$}  & \multirow{2}{*}{$B_2 {( N + 2 q+ 1)}$}\\
			$w_+^+$&$(2 n_2 -1 ) B_2$ & & \\
\rule{0pt}{4ex}
$W_-^+$& {$(2 n_2 +1 )B_2$ } &  $\hspace{-1em}~+~ \bigl( q^2 + q (N+3) + 3 (N/2) + 2\bigr) /a^2$&$B_2 {( N + 2 q +3)} $\\
\rule{0pt}{4ex}
&\multirow{3}{*}{$(2 n_2 +1 )B_2$ } & $ \hspace{-1em}~+~\bigl( q^2 + q(N-1) - (N/2) \bigr)/a^2$ & $B_2 {( N + 2 q- 1)} $, ~{\rm if}~~$N\geq 2$\\
$W_+^+$& &$\hspace{-1em}~+~(q^2 + 2 q+ {\half} )/a^2$& $B_2 (2 q +2)$, ~~~~~~~{\rm if}~~$N=1$\\
&&$\hspace{-1em}~+~(q^2 + 3 q+ 2 )/a^2$&$B_2 (2 q + 3)$, ~~~~~~~{\rm if}~~$N= 0$\\
\bottomrule[\heavyrulewidth] 
		\end{tabular}
	\end{center}
\end{table} 
From the values given in the table, there is a useful observation we can make regarding
the decay rate. Notice that $w_+^+, \, w_-^+$ have the same eigenvalues and degeneracies
as the ghosts except for an additional $\mp2 B_2$, respectively.
In the formula for the effective action, this will give an additional factor
$e^{-s (\mp 2 B_2)}$. Upon continuation to Minkowski space, when this is evaluated at the
poles of the $(\sin sE)^{-1}$  factor, we get $e^{\mp 2\pi i n} = 1$. So these factors will not
affect the decay rate. Since all other terms are identical, the contribution of
the $w$-fields (i.e., $W_3, \, W_4$) is exactly canceled by the ghosts.
Again, this is essentially the reduction of the degrees of freedom to the two
physical polarizations. The decay rate can thus be obtained from just the
$W_\pm^\pm$-fields and is given by
\be
\Ree (i S_{\rm eff}) = - \int d\mu dx_0 dx_3 {E^2 \over 16 \pi^3}
\, \beta_1(\omega) \,, \label{Sp-vec3a}
\ee
\begin{multline}
\beta_1(\omega) = \omega \sum_{n=1}^\infty
{(-1)^{n+1} \over n}\sum_q \Bigl[  (2 q +N +3 ) e^{- n\omega \bigl(q^2 + q (N+3) + 3(N/2) +2 +m^2a^2\bigr)} \\
+ (2 q +N -1 ) e^{- n\omega \bigl(q^2 + q (N-1) - (N/2) +m^2a^2\bigr)}\Bigr]\, ,
\label{Sp-vec4}
\end{multline}
where $\omega = \pi /(Ea^2)$.
For $N\geq 2$, the eigenvalue for $W_+^+$ for $q =0$ is negative, namely $-(N/2)$, so there is
a convergence problem for the summation over $n$. So we have again added
a mass term as an ad hoc infrared regulator.
The eigenvalues and degeneracy for $W_+^+$ for $q \geq 2$ coincide
with those of $W_-^+$. Thus after separating out the $q =0, 1$ terms
for the $W_+^+$ spectrum and redefining $q$,
we can write
the formula for $\beta_1 (\omega)$ as
\beqar
\beta_1 (\omega ) &=& \omega \sum_{n=1}^\infty {(-1)^{n+1} \over n} \Bigl[ (N-1) e^{-n \omega (m^2 a^2 - (N/2))}
+ (N+1) e^{- n \omega (m^2 a^2+ (N/2))} +\nonumber\\
&&\hskip 1in + 2 \sum_{q =1}^\infty (2 q +N +1) e^{ - n \omega (m^2 a^2 + q^2 + q (N+1) + (N/2))}
\Bigr]\nonumber\\
&=&\omega \Bigl[ (N-1) \log \bigl(1 + e^{-\omega (m^2 a^2 - (N/2))}\bigr)
+ (N+1) \log \bigl(1 + e^{-\omega (m^2 a^2+ (N/2))}\bigr)\nonumber\\
&&\hskip 1in 
+ 2 \sum_{q=1}^\infty (2 q + N +1 ) \log \bigl( 1+ e^{- \omega (m^2 a^2+ q^2 + q (N+1) +(N/2))}\bigr)
\Bigr]
\label{Sp-vec5}
\eeqar
This equation holds for any $N \geq 1$.
It is easy to verify that, for large $a^2$ or the flat limit of $S^2$, we can approximate
$\beta_1 (\omega )$ by
\beqar
\beta^{\rm flat}_1(\omega )
&=&  \omega \sum_{n=1}^\infty
{(-1)^{n+1} \over n} N \sum_q \Bigl[  e^{- n\omega \bigl( q N + 3(N/2) +m^2 a^2 \bigr)}
 +  e^{- n\omega \bigl(q N - (N/2) +m^2 a^2 \bigr)}\Bigr]\nonumber\\
 &=&\omega N \Bigl[ \log \bigl( 1 + e^{-\omega (m^2 a^2 - (N/2))}\bigr)
 + \log \bigl( 1 + e^{-\omega (m^2 a^2 + (N/2))}\bigr) \nonumber\\
 &&\hskip .4in + 2 \sum_{q=1}^\infty
 \log \bigl( 1+ e^{- \omega (m^2 a^2 + q N + (N/2))}\bigr)\Bigr]
 \label{Sp-vec6}
 \eeqar
 If we use this limiting value in (\ref{Sp-vec3a}) and take $m = 0$, we get the formulae
 (\ref{rev27}), (\ref{rev28})
which apply for the flat case with a background field, with the identification
$N = 2 B_1 a^2$.
 We can now define
 \beq
 \gamma_1 (\omega ) = \left[ { \beta_1 (\omega ) \over \beta^{\rm flat}_1(\omega)}
 \right]
 \label{Sp-vec7}
 \eeq
This gives a good measure of the effect of curvature.
In the absence of the transverse magnetic field, i.e., for the $N=0$ case, $\beta_1 (\omega)$ takes the form
\be
\beta_1 (\omega, N=0) = 2 \omega \sum_{q=1}^{\infty} (2 q +1) \log (1 + e^{- \omega q(q+1)})
\label{beta1sphere}
\ee
while the corresponding quantity in the flat limit is ${\pi^2/6}$. Therefore, for $N =0$, we find
$\gamma_1 (\omega, N=0) = \frac{6}{\pi^2}\beta_1 (\omega, N=0)$. 
The graphs of $\gamma_1 (\omega )$, for different values
of the magnetic field, or $N$, are shown in Figs.\,\ref{gamma_1-1} and \ref{gamma_1-2}.
It is clear that the main features of these graphs are independent of the cut-off used.
Notice that, for all values of $N$ ($\neq 0$), $\gamma_1(\omega)$ is less than $1$, with an asymptotic limit
of $(N-1)/N$
for large $\omega$.
Thus the 
effect of the positive curvature of $S^2$ is to suppress pair production compared to
the rate in flat space. As the asymptotic value shows, this effect is essentially 
due to the degeneracy
factors. (In broad terms, the situation is similar to what happens for
spin-$\half$ fields, but is very different from the result for
scalar fields, see \cite{KKN1}.)
It is also clear that the pair production rate is higher with a background
magnetic field
than it is for zero magnetic field
since the graphs show clearly that $\gamma_1 (\omega, N ) > \gamma_1 (\omega, N = 0)$,
although the enhancement is less pronounced than it is for the flat case, since $\gamma_1 (\omega, N )
< 1$.

\begin{figure}[!htb]
		\centering
\begin{minipage}[t]{0.5\textwidth}
 		\centering
		\includegraphics[width=0.9\textwidth]{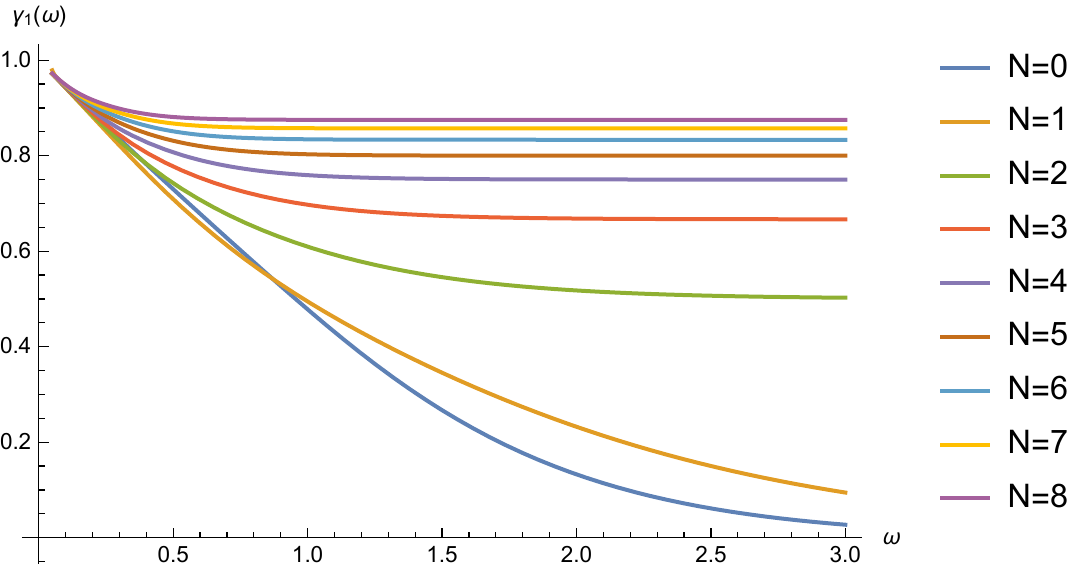}
		\caption{ Cut-off $m^2 a^2 = 4$} 
		\label{gamma_1-1}
 	\end{minipage}%
 	\begin{minipage}[t]{0.5\textwidth}
 		\centering	
	\includegraphics[width=0.9\textwidth]{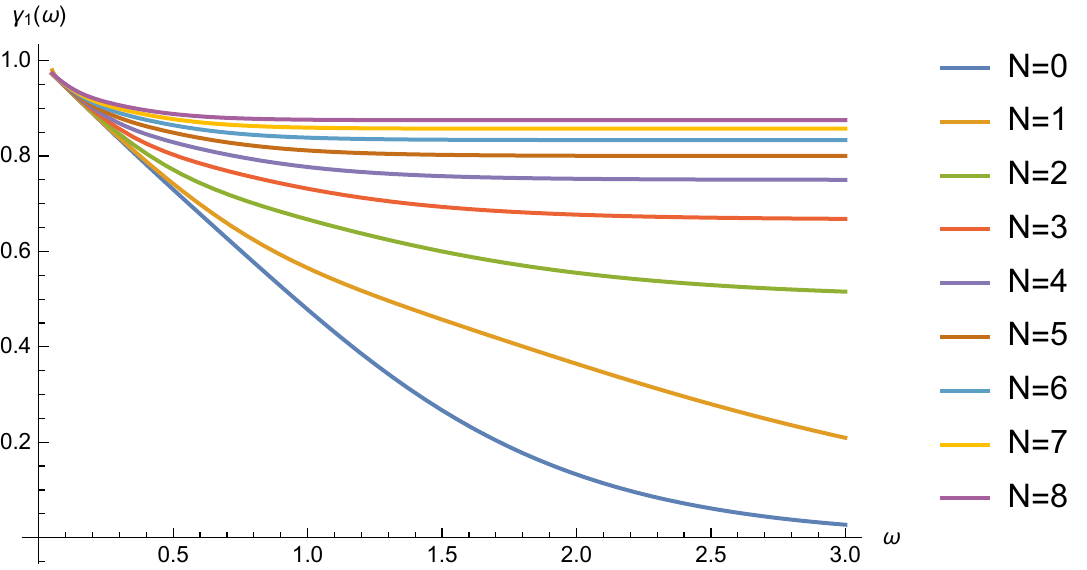}
	\caption{No cut-off, $m^2 a^2 =0$} 
	\label{gamma_1-2}
	 	\end{minipage}
\end{figure}
In this example of $S^2 \times {\mathbb R}^{1,1}$ also, it is instructive to take the
small $E$ limit, or large $\omega$ limit. The expression for the real part of
$i S_{\rm eff}$ becomes
\beq
\Ree (i S_{\rm eff}) = - \int d\mu dx_0 dx_3 \,
{1\over 8 \pi}\, B_1\, \left( B_1 - {1\over 2 a^2}\right) =
 - \int d\mu dx_0 dx_3 \,
{1\over 8 \pi}\, B_1\, \left( B_1 - {R\over 4}\right)
\label{Sp-vec8}
\eeq
where $R$ is the Ricci scalar for the sphere, $R = 2/a^2$. Notice that the instability is
cured by a small enough radius for the sphere, $ 2 B_1 a^2 = 1$.  This is intuitively in agreement
with curing the instability with a mass term \cite{mass}; both provide
suitable infrared cutoffs.

We also want to contrast this result with the calculation reported in
\cite{elizalde}, where the imaginary part of the effective action is obtained as
\beq
\Ree (i S_{\rm eff}) = - \int d\mu d^2x
{1 \over 8 \pi} \,B_1\,\left( B_1 + {1\over 3 a^2}\right)
\label{Sp-vec9}
\eeq
(We have rewritten the formula in our notation.)
It should be emphasized that the case considered in \cite{elizalde}
is for $S^2 \times {\mathbb R}^2$ with the magnetic field
purely in the ${\mathbb R}^2$ part and not on the $S^2$, as we are doing here.
So, while (\ref{Sp-vec9}) is an interesting result, it cannot be compared to
our calculation.

\section{Pair Creation of Vector Particles on $H^2 \times {\mathbb R}^{1,1}$}

As before we consider a magnetic field $B_1$ on $H^2$ and a magnetic field
$B_2$ on ${\mathbb R}^2$ (which will be continued to the electric field on
${\mathbb R}^{1,1}$). We also define $b = \vert B_1 \vert a^2$, where
 the curvature on $H^2$ is $-2/a^2$.
The space $H^2$ can be analyzed using group theory in a way similar to what we did
for $S^2$, since $H^2 = {SU(1,1)}/{U(1)}$.
The eigenvalues for the Laplacian for $H^2$ can be obtained in
terms of unitary representations of $SU(1,1)\sim SL(2, {\mathbb R})$, as explained
in \cite{{Comtet:1984mm},{Comtet:1986ki}} and \cite{KKN1}.
The generators $R_\pm$, $R_3$ of $SL(2, {\mathbb R})$ satisfy the commutation relations
\be
\lbrack R_3 \,, R_\pm \rbrack =  \pm R_\pm \,, \quad \lbrack R_+ \,, R_- \rbrack = -2 R_3 \,.
\label{su11}
\ee
The wave functions are the group elements of $SL(2, {\mathbb R})$
with $R_3$-values fixed by the magnetic field and
the curvature, similar to what we did for $S^2= SU(2)/U(1)$.
The relevant representations in the
present case are the 
the discrete series bounded below and 
the principal continuous series.
In considering the kinetic operators, which have the spin-magnetic field and 
spin-curvature couplings as well, we note
that, since $H^2$ has constant negative curvature, the sign of the curvature term used in the spherical case (\ref{Sp-vec3})  now flips to the negative sign. Therefore, we have
\be
- \Tr W_\nu [D_\mu, D_\nu]W_\mu = - {1\over 2 a^2} \bigl( W^c_- R_3 W^c_+ - W^c_+ R_3 W^c_-)
= - {1\over a^2} \, W^c_- W^c_+
\label{H2-curvature term}
\ee
Keeping these facts in mind, the energy eigenvalues and the corresponding densities for the charged fields can be determined in a straightforward manner. 
The eigenvalues of the $H^2$ part of the kinetic operator
for $W_\pm^+$, including the spin-magnetic field and spin-curvature terms,
are of the form
\beq
-{\cal D}_{W_\pm^+} = {1\over a^2}
\left[ (R^2 +R_3^2) \mp 2 b - 1\right]
\label{H2-eigen}
\eeq
\begin{table}[!b]
	\caption{Charged Fields and the Eigenvalues for the Continuous part of the Spectrum on $H^2$}  
	\label{cont-eigen}
	\small 
	\centering 
	\begin{tabular}{ccl c} 
		\toprule[\heavyrulewidth]\toprule[\heavyrulewidth]
		\textbf{Field} & \textbf{Eigenvalues} & \textbf{} & \textbf{$\bm {4 \pi^2 a^2 \, \rho}$} \\ 
		\midrule
$c^\pm$&$(2 n_2 +1) B_2$&{$ \hspace{-1em} + \bigl(\lambda^2 + \frac{1}{4} + b^2\bigr)/a^2$} &{$B_2 \frac{ \lambda \, \sinh 2 \pi \lambda}{\cosh 2\pi \lambda + \cos 2 \pi b}$} \\
\rule{0pt}{4ex}
		$w_-^+$& $(2 n_2 +3) B_2$ & \multirow{2}{*}{$ \hspace{-1em} + \bigl(\lambda^2 + \frac{1}{4} + b^2\bigr)/a^2$} &\multirow{2}{*}{$B_2 \frac{ \lambda \, \sinh 2 \pi \lambda}{\cosh 2\pi \lambda + \cos 2 \pi b}$} \\
		$w_+^+$&$(2 n_2 -1 ) B_2$ \\
\rule{0pt}{4ex}
		$W_-^+$& \multirow{2}{*}{$(2 n_2 +1 )B_2$}& \multirow{2}{*}{$ \hspace{-1em} + \bigl(\lambda^2 + \frac{1}{4} + b^2\bigr)/a^2$} &\multirow{2}{*}{$B_2 \frac{ \lambda \, \sinh 2 \pi \lambda}{\cosh 2\pi \lambda + \cos 2 \pi b}$} \\
		$W_+^+$& \\ 

		\bottomrule[\heavyrulewidth] 
	\end{tabular}
\end{table}
where $R_3 = b \pm 1$ for $W_\pm^+$ and $R^2$ is the eigenvalue for the quadratic Casimir operator
of $SU(1,1)$. Explicitly, for the principal continuous series representation,
$R^2 = \lambda^2 + {1\over 4}$, where $\lambda$ is real, $0\leq \lambda < \infty$
and for the discrete series $R^2 = - \Lambda (\Lambda -1)$ where
$\Lambda = R_3 -k \geq {1\over 2}$. We use the discrete series representations
which are bounded
below as is appropriate with the finite norm condition defined by the
parametrization we have chosen for $H^2$ \cite{KKN1}.
Detailed calculations are given in the appendix A and the results are summarized in the tables given below for convenience.
Table\,\ref{cont-eigen} gives the continuous part of the energy spectrum on $H^2$, while 
Table\,\ref{disc-eigen} refers to the discrete part.
\begin{table}[!t]
	\caption{Charged Fields and the Eigenvalues for the Discrete part of the Spectrum on $H^2$}  
	\label{disc-eigen}
	\small 
	\centering 
	\begin{tabular}{ccl c} 
		\toprule[\heavyrulewidth]\toprule[\heavyrulewidth]
		\textbf{Field} & \textbf{Eigenvalues} & \textbf{} & \textbf{$\bm {4 \pi^2 a^2 \, \rho}$} \\ 
		\midrule
$c^\pm$& $(2 n_2 +1 ) B_2$ &{$ \hspace{-1em} + \bigl(-k(k+1) +2 b k + b \bigr)/a^2 \,, \quad \quad \quad ~k \leq \lbrack b-\frac{1}{2} \rbrack$} &~{$B_2 (b - k - \frac{1}{2})$} \\
\rule{0pt}{4ex}
		$w_-^+$& $(2 n_2+3) B_2$ & \multirow{2}{*}{$ \hspace{-1em} + \bigl(-k(k+1) +2 b k + b \bigr)/a^2 \,, \quad \quad \quad ~k \leq \lbrack b-\frac{1}{2} \rbrack$} &\multirow{2}{*}{~$B_2 (b - k- \frac{1}{2})$} \\
		$w_+^+$&$(2 n_2 -1 ) B_2$ \\
\rule{0pt}{4ex}
		$W_-^+$&\multirow{2}{*}{$(2 n_2 +1 )B_2$}& $\hspace{-1em} + \bigl(-k(k+3) +2 b k + 3 b -2 \bigr)/a^2 \,, \quad k \leq \lbrack b - \frac{3}{2} \rbrack$ & ~$B_2 (b - k- \frac{3}{2})$ \\ 
		$W_+^+$& & $\hspace{-1em} + \bigl(-k(k-1) +2 b k - b \bigr)/a^2 \,, \quad \quad
		\quad ~k \leq \lbrack b + \frac{1}{2} \rbrack$ & ~$B_2 (b - k + \frac{1}{2})$ \\

		\bottomrule[\heavyrulewidth] 
	\end{tabular}
\end{table}
The discrete energy levels are labeled by an integer $k \geq 0$.
It is clear from the table of discrete levels that, for $0 \leq b <\half$ only a single discrete energy level exists for $W_+^+$, which has energy $-b/a^2$ on $H^2$, which is a zero mode in the absence of the magnetic field. For $\frac{1}{2} < b < \frac{3}{2}$, there is an additional
discrete level for $W_+^+$ with energy $b/a^2$. For $b> \frac{3}{2}$, there is a finite number
of discrete states, labeled by the
integer $k \geq 0$ with $k \leq \lbrack b-\frac{3}{2} \rbrack$.  
($\lbrack  X \rbrack$ indicates the integer part of the argument $X$.)
It is also easily verified that the continuum starts at a higher energy
than the highest discrete level.

Contributions to the imaginary part of the effective action coming from the $w_\pm^\pm$ and the ghosts $c^\pm$, cancel in the same manner as they do in the $S^2 \times {\mathbb R}^{1,1}$ case. This leaves us with the contributions coming from the $W_\pm^\pm$ fields, which lead to 
\be
{\Ree}(i S_{\rm eff}) =
- \int_{H^2\times {\mathbb R}^{1,1}} d\mu d x_0 \, d x_3\, \frac{E^2}{16 \pi^3}  \beta_1 (\omega)
\label{SEFFH2M2}
\ee
where $\beta_1(\omega) = \beta_{1,C}(\omega) + \beta_{1,D}(\omega)$, with 
\be
\beta_{1,C}(\omega) = 4 \omega \int_0^\infty d \lambda \, \frac{ \lambda \, \sinh 2 \pi \lambda}{\cosh 2\pi \lambda +  \cos 2 \pi b} 
\log [1 + e^{-\omega(\lambda^2 + \frac{1}{4} + b^2 + m^2 a^2)}] 
\label{beta-cont}
\ee
\begin{multline}
\beta_{1,D}(\omega) = 2 \omega \sum_{n=1}^\infty \frac{(-1)^{n+1}}{n} \Bigg ( \sum_{k=0}^{\lbrack b+
{\textstyle{1\over 2}} \rbrack} (b-k+\frac{1}{2}) e^{- \omega n \big(-k(k-1) +2 b k - b + m^2 a^2 \big)} \\
+ \sum_{k=0}^{\lbrack b- {\textstyle{3\over 2}} \rbrack} (b-k - \frac{3}{2}) e^{- \omega n \big (-k(k+3) +2 b k + 3 b -2 + m^2 a^2 \big)} \Bigg )
\label{beta-cont2}
\end{multline}
After separating out the $k=0,1$ terms, the first sum in (\ref{beta-cont2}) gives an identical contribution to the second and we can express, $\beta_{1,D}(\omega)$ as
\beqa
\beta_{1,D}(\omega) &=& 2 \omega \Bigg \lbrack (b + \frac{1}{2}) \log(1+e^{-\omega (-b + m^2 a^2)}) 
+ (b - \frac{1}{2}) \log(1+e^{-\omega (b+ m^2 a^2)}) \nn \\
&& \quad \quad + 2 \sum_{k=0}^{[b-{\textstyle{3\over 2}}]} (b-k-\frac{3}{2}) \log (1+e^{-\omega (-k(k+3)+2 b k + 3 b -2 + m^2 a^2)}) \Bigg \rbrack \label{beta-disc}\\
&\equiv& \beta_{1,D}^{(<1/2)}(\omega) + \beta_{1,D}^{(<3/2)}(\omega) + \beta_{1,D}^{(>3/2)}(\omega) \,, \nn
\eeqa
where in the last line  we have introduced a helpful notation to facilitate the fact that not all the 
terms in $\beta_{1,D}(\omega)$ are present for all
$b$.
Thus, only $\beta_{1,D}^{(<1/2)}(\omega)$ is present for $0 \leq b < \frac{1}{2}$, $\beta_{1,D}^{(<1/2)}(\omega)+\beta_{1,D}^{(<3/2)}(\omega)$ for $\half < b < \frac{3}{2}$ and all three terms are present 
for $b > \frac{3}{2}$. The degeneracy factors should be positive; this gives a quick
check on which terms are present when.

In the flat limit of the hyperbolic plane, $\beta_{1,C}(\omega)$ does not give any contribution, 
while $\beta_{1,D}(\omega)$ takes the form
\beqar
\beta^{\rm flat}_{1,D}(\omega) &=& 2 \omega b \sum_{k=0}^{k_{max} \rightarrow \infty} \Big  \lbrack \log(1+e^{-\omega (2 b k -b + m^2 a^2)}) + \log(1+e^{-\omega (2 b k +3 b + m^2 a^2)}) \Big \rbrack 
\nonumber\\
&=&2 \omega b \Bigg ( \log(1+e^{-\omega (-b + m^2 a^2)}) + \log(1+e^{-\omega (b + m^2 a^2)}) \nonumber\\
&&\hskip .4in + 2 \sum_{k=0}^{k_{max} \rightarrow \infty}
\log(1+e^{-\omega (2 b k +3 b + m^2 a^2)}) \Bigg) \,.
\eeqar
As before, to probe the curvature effects at a given magnetic field, 
we compare $\beta_1(\omega)$ to its flat space value by defining
the functions
\beqar
\gamma_1 (\omega) =
\begin{cases} 
\frac{\beta_{1,D}^{(<1/2)}(\omega)+\beta_{1,C}(\omega)}{\beta^{\rm flat}_{1,D}(\omega)}\, , & \hskip .4in  0\leq b < \half  \\
\frac{\beta_{1,D}^{(<1/2)}(\omega)+\beta_{1,D}^{(<3/2)}(\omega)+\beta_{1,C}(\omega)}{\beta^{\rm flat}_{1,D}(\omega)}\, , & \hskip .4in \half < b < {\textstyle{3\over 2}} \label{gammas}\\
\frac{\beta_{1,D}(\omega)+\beta_{1,C}(\omega)}{\beta^{\rm flat}_{1,D}(\omega )} \, , & \hskip .6in
b> {\textstyle{3\over 2}} \,.
\end{cases}
\eeqar
For the special case of $b=0$, we find
\be
\gamma_1^{(0)}(\omega) = \frac{6}{\pi^2} \left (  \omega \log 2 + 4\, \omega \int_0^\infty d \lambda \, \lambda \tanh \pi \lambda \log \lbrack 1+ e^{- \omega (\lambda^2 + \frac{1}{4})} \rbrack \right ) \,,
\ee

In the absence of the transverse magnetic field, in addition to the contribution from the continuous energy spectrum, there is a single discrete mode, which is a zero mode, with constant density of states, whose contribution to $\gamma_1^{(0)}(\omega)$ is $\frac{6}{\pi^2} \omega \log 2$. This mode is 
essentially responsible for the monotonic increase in the pair production rate compared to the flat case by accommodating produced particles at virtually no energy cost. This feature
is clearly seen in the profile of $\gamma_1^{(0)}(\omega)$ provided in Fig.\,\ref{H2M2_Vector_Fig1}.

In order to understand the emerging physics from the profiles of $\gamma_1(\omega)$,
several facts should be simultaneously taken into account. From Figs.\,\ref{H2xM2VectorFig8}-\ref{H2xM2VectorFig3}, we can see that the pair production rate on $H^2 \times {\mathbb R}^{1,1}$ is always larger than that for vector particles on flat space and converges to the latter at sufficiently large magnetic fields. Next, it is important to emphasize that all the ensuing results regarding the comparison of pair production rates at different value of the magnetic field or comparison with the results obtained in the flat case are independent of the value of the mass term $m^2 a^2$, which is acting as an effective infrared cut-off. In retrospect, this may be expected since the infrared cut-off is introduced formally to 
facilitate certain summations because the
negative energy mode is present in the spectrum of $W_+^+$. As such, the cut-off appears in all the exponentials in the expressions for $\gamma_1(\omega)$, and that makes the latter almost completely insensitive to whatever value it may take ( as long as it is not unphysically large). From now on, we therefore set it to zero without loss of generality.     

For $0 < b < \frac{1}{2}$, we observe from the profiles of $\gamma_1(\omega)$ in 
Fig.\,\ref{H2xM2VectorFig8} that, there is further increase in the pair production effect over and above the rate at $b=0$. For this range of values for the magnetic field there is still just one discrete mode but now with energy $- \frac{b}{a^2} + m^2 $, which, in the absence of infrared cut-off $m^2$, is the one and only negative energy mode. This, in itself, is sufficient to render the effect larger than what it is
for the flat case and also larger than for the
$b=0$ case as long as $\omega$ is not too large. The reason for this is the larger degeneracy of this state compared to that of the corresponding state in the flat limit; i.e.,
 $b + \frac{1}{2} > b$, with the extra $\half$ due to the non-zero curvature. In fact, for sufficiently large $\omega$, we infer that $\gamma_1(\omega)$ goes like $\simeq 1 + \frac{1}{2b} > 1$.
 

For $\half < b < \frac{3}{2}$, there are two discrete modes, one with
energy $-b/a^2$ and  one with energy $b/a^2$. Profiles 
of $\gamma_1(\omega)$ in Fig.\,\ref{H2xM2VectorFig7} show that the enhancement in pair production over and above the $b=0$ case is sustained in a shorter interval of $\omega$ which gets gradually narrower with increasing $b$-field. The observed behavior of $\gamma_1(\omega)$ can be anticipated from the foregoing discussion, since, besides the opposing effect from the continuous energy levels, the additional discrete level also becomes costlier to fill with increasing $b$.

Finally, for $b > \frac{3}{2}$, there are as many additional discrete energy levels as
is consistent with $q <[b-3/2]$. From the profiles of $\gamma_1(\omega)$
shown in Figs.\,\ref{H2xM2VectorFig2}, \ref{H2xM2VectorFig3}, we see that the influence of increasing magnetic field on pair production rate is to drive its value back towards
that for the flat case, since it becomes progressively costlier in energy for produced particles to fill the available quantum states at higher magnetic fields. (We include two sets of figures to emphasize that the basic features are
independent of the cut-off.)
\begin{figure}[!t]
	\centering
	\includegraphics[width=0.5\textwidth,height=0.2\textheight]{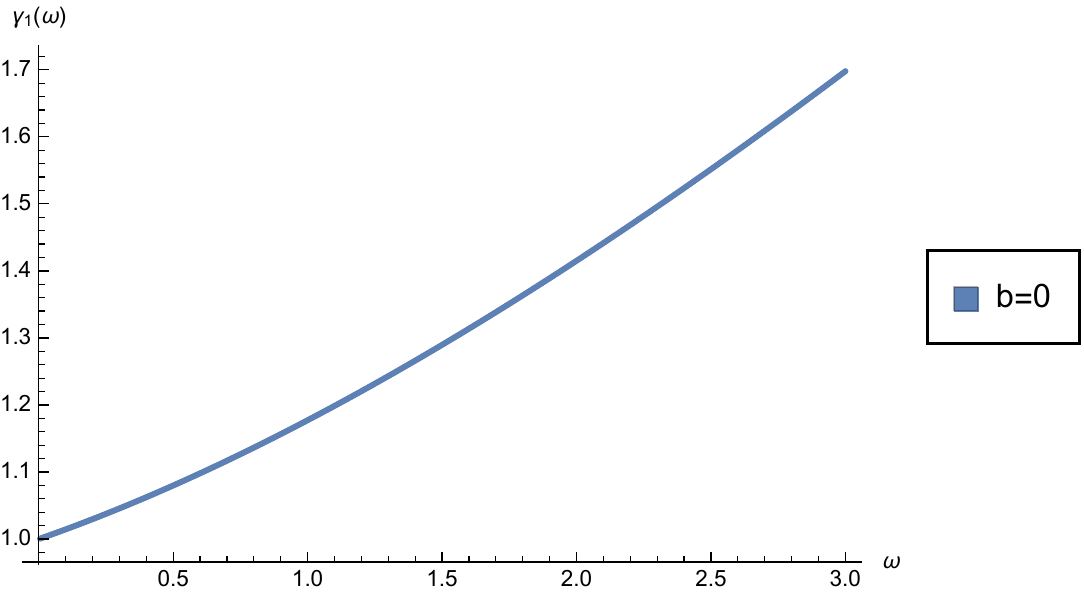}
	\caption{$\gamma_1(\omega)$ at $b=0$.}
	\label{H2M2_Vector_Fig1}
\end{figure}
\begin{figure}[!htb]\centering
	\begin{minipage}[t]{0.47\textwidth}
		\centering
		\includegraphics[width=1\textwidth]{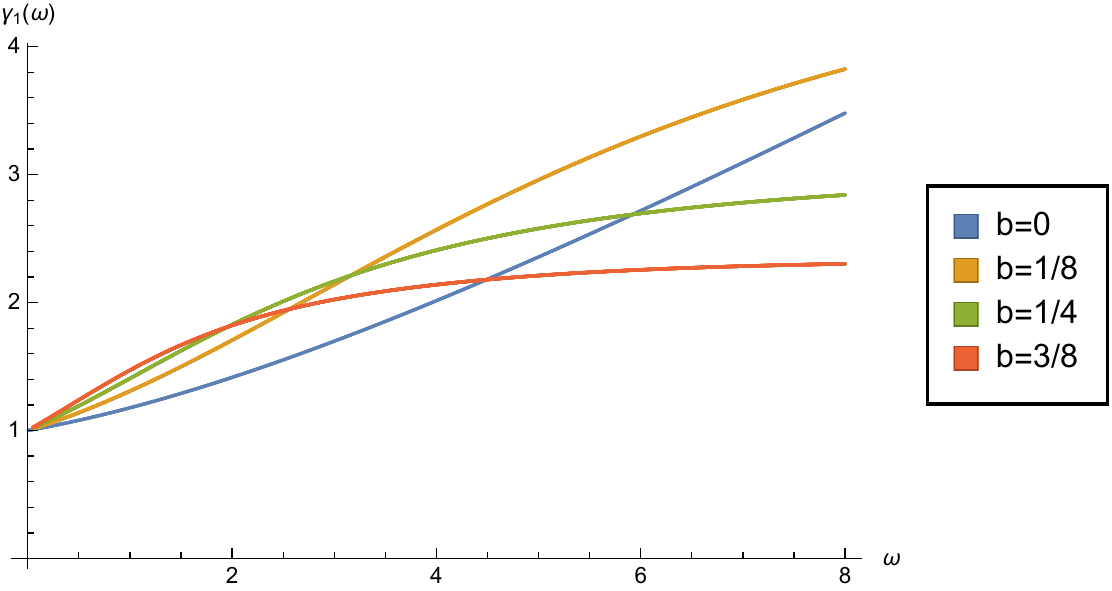}
		\caption{$\gamma_1(\omega)$ at $ 0 < b < \frac{1}{2}$ and no infrared cut-off.} 
		\label{H2xM2VectorFig8}
	\end{minipage}%
	\hskip .1in
	\begin{minipage}[t]{0.47\textwidth}
		\centering	
		\includegraphics[width=1\textwidth]{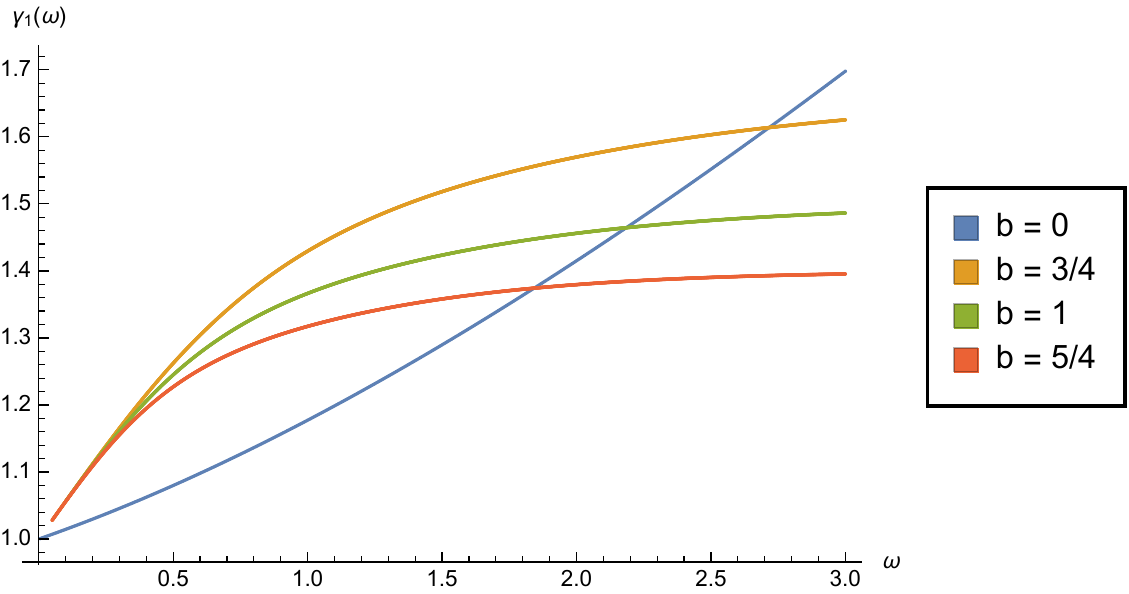}
		\caption{$\gamma_1(\omega)$ at $ \frac{1}{2} < b < \frac{3}{2}$ and no infrared cut-off.} 
		\label{H2xM2VectorFig7}
	\end{minipage}
\end{figure}

\begin{figure}[!htb]\centering
	\begin{minipage}[t]{0.47\textwidth}
		\centering
		\includegraphics[width=1\textwidth]{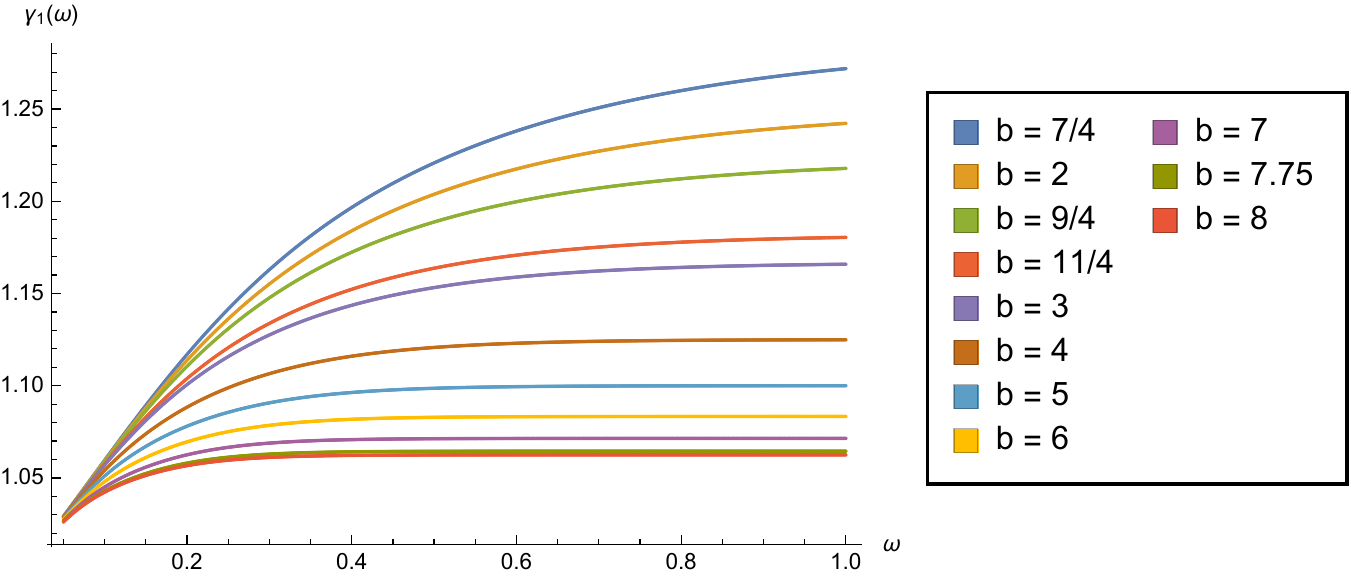}
		\caption{$\gamma_1(\omega)$ at $ b > \frac{3}{2}$ and infrared cut-off value $m^2a^2 = 8  $.} 
		\label{H2xM2VectorFig2}
	\end{minipage}%
	\hskip .1in
	\begin{minipage}[t]{0.47\textwidth}
		\centering	
		\includegraphics[width=1\textwidth]{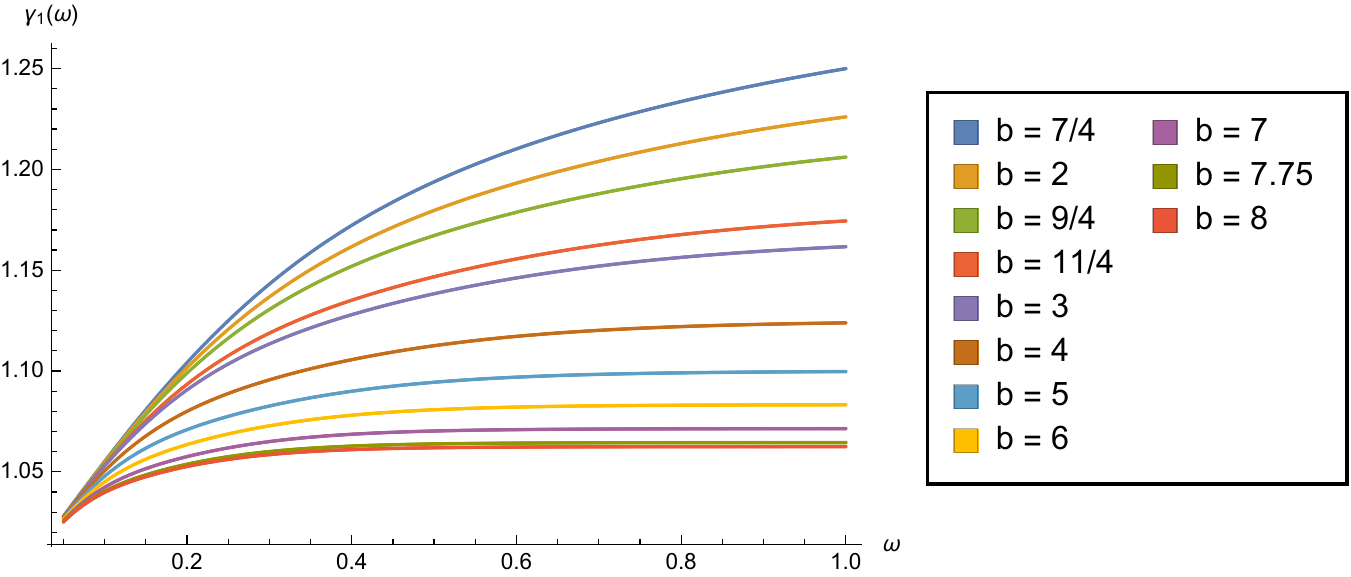}
		\caption{$\gamma_1(\omega)$ at $ b > \frac{3}{2}$ and no infrared cut-off.} 
		\label{H2xM2VectorFig3}
	\end{minipage}
\end{figure}
Finally, as in the case of flat space and $S^2 \times {\mathbb R}^{1,1}$, we can take the limit of
$E \rightarrow 0$ or $\omega \rightarrow \infty$. From equations
(\ref{beta-cont}) and (\ref{beta-disc}), we see that only 
$\beta_{1,D}^{(<1/2)}(\omega)$ can give a nonzero value in this limit.
The result is that
\beq
{\Ree}(i S_{\rm eff}) =
- \int d\mu d x_0 \, d x_3\, \frac{1}{8 \pi} B_1 \left( B_1 + {1\over 2 a^2}
\right) =
- \int d\mu d x_0 \, d x_3\, \frac{1}{8 \pi} B_1 \left( B_1 - {R\over 4}
\right)
\label{instab-H2}
\eeq
Comparison of this formula with (\ref{rev28a}) and (\ref{Sp-vec8}) shows
that this formula, as written in terms of the Ricci scalar, captures the general result for
the Nielsen-Olesen instability for
all three cases, $M= {\mathbb R}^2$, $S^2$, $H^2$.

\section{Summary and remarks on QCD vacuum and confinement}

We calculated the pair production rate for vector particles, 
and corresponding decay of a background
electric field, on manifolds of the form $M \times {\mathbb R}^{1,1}$,  with a background magnetic field
on $M$, where
$M= {\mathbb R}^2$, $S^2$ and $H^2$. In order for this to be embedded in a consistent theory
of vector particles we used the Yang-Mills action. The latter specified the spin-magnetic field
and spin-curvature couplings. The pair production rate is enhanced by the negative eigenvalues
for the kinetic operator due to the spin-magnetic field coupling.
The additional spin-curvature coupling suppresses this effect to some extent
for positive curvature
and enhances it further for negative curvature. Comparison of $\gamma_1 (\omega)$ for $S^2$ and $H^2$ at zero magnetic field can be made by inspecting Fig.\,\ref{gamma_1N=0},
which shows the deviation of
$\gamma_1(\omega)$ for $S^2$ and $H^2$ compared to the flat case
which has $f_1(x=0) = 1$ for $M={\mathbb R}^2$ for all $\omega$.
\begin{figure}[!t]
	\centering
	\scalebox{.8}{\includegraphics[width=0.6\linewidth,height=0.25\textheight]{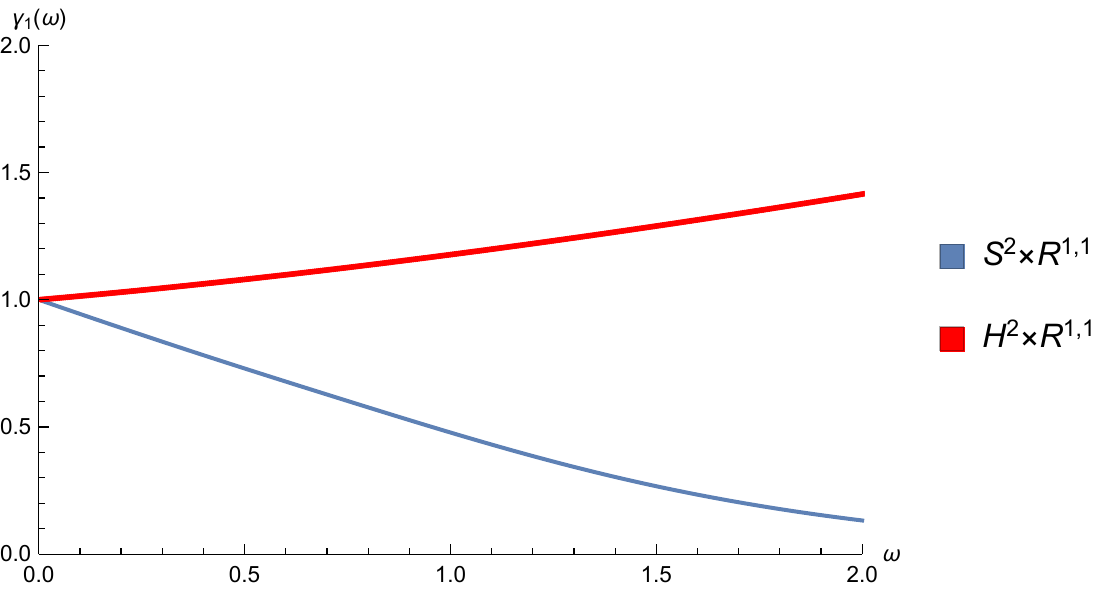}}
	\caption{Comparison of $\gamma_{1}$ at zero magnetic field for the sphere and hyperboloid}
	\label{gamma_1N=0}
\end{figure}
Our calculations also give the
generalization of the
Nielsen-Olesen instability to include nonzero curvature.

Since we obtained the action for charged vector particles by considering an expansion around a background field of the standard Yang-Mills action, our calculations have some
implications for nonabelian gauge theories. Admittedly, even though our calculations are not entirely perturbative, they are
still equivalent to a one-loop effective action and hence it is not possible to
make definitive conclusions about
confinement and related phenomena. Nevertheless, we know that
there are many calculations, which, while
not definitive, do carry
intimations of confinement. Beyond the well-known issue of asymptotic freedom, among these
we can include the difficulty with unitary implementation of color rotations \cite{Bal},
the problem of sustaining a statistical distribution of nonzero color charge \cite{NY}, etc.
Our calculation of the pair production in this paper, coupled with the known instability of chromomagnetic fields,
leads to a similar suggestive view on confinement.

A consequence of asymptotic freedom is that the vacuum of QCD has a tendency to spontaneously develop nonzero expectation values
for chromomagnetic fields. In other words, a state with a nonzero magnetic field can have
lower energy than the perturbative vacuum with zero field. This was noticed decades ago
\cite{savv}.
It has been used as the basis for assigning a nonzero vacuum value
$\bra{0} F^2 \ket{0}$, and used in sum rules \cite{F^2}.
A chromomagnetic field, however,  can lead to instability due to the negative eigenvalue for
the kinetic operator arising from the spin-magnetic field coupling \cite{NO}.
This is also clear from our equations
(\ref{rev28a}), (\ref{Sp-vec8}), (\ref{instab-H2}).
There have been many attempts to
use this observation due to Nielsen and Olesen to develop an understanding
of the nonperturbative confining vacuum of Yang-Mills theory, leading to the so-called 
spaghetti vacuum,
or Copenhagen vacuum \cite{NO}.
Arguments have also been made that the instability could be cured by a new vacuum state which generates
a ``mass" for the gluons \cite{mass}.

Combining these observations with the calculations in this paper
gives another perspective on some aspects of
confinement. Asymptotic freedom 
moves the theory in the direction of generating a chromomagnetic field. 
Such a field, by our arguments,
leads to a highly enhanced decay rate for any chromoelectric field. Our 
calculations are for uniform fields,
but they should still apply approximately to fields which are uniform over 
some small range. Thus any chromoelectric field decays at an enhanced rate
by pair production. But
the particles produced in this process of the decay of the field
are themselves charged and have their own chromoelectric fields, so in principle, the process
can continue.
(For the electromagnetic case, a similar statement is true as well, but the field carried by the decay products are weak and further pair production is suppressed by mass thresholds.)
Thus further decays (of the chromoelectric fields of the
produced pairs), in fact a whole cascade of decays, can be terminated and 
stability obtained only if
the charged particles which are produced combine into color-singlets and
so become free of any accompanying chromoelectric fields.
This gives a dynamic view of how confinement could arise.

Admittedly, the calculations we have given have limited validity. But we see that, even within
a one-loop background field approximation, or within a resummation of one-loop
diagrams, there are serious difficulties in maintaining chromoelectric fields.

\appendices

\section{Spectrum of the kinetic operator for vectors on $H^2$}
\setcounter{equation}{0}

Using (\ref{rev25}) and (\ref{H2-curvature term}) it is straightforward to see that the relevant quadratic differential operators for $W_\pm^+$ on $H^2$ are given as
\be
- {\cal D}^2_{V} := - D^2_{H^2} \mp 2 \frac{b}{a^2} - \frac{1}{a^2} \,,
\label{DVector1}
\ee
where the second and third terms in the right hand side
 are due to the spin-magnetic field and the spin-curvature couplings, respectively. 

Following the discussion and the results given in the appendix of \cite{KKN1}, we can determine the discrete and the continuous spectrum of $- {\cal D}^2_{V}$. 
(For general references on the relevant representation theory, see
\cite{Perelomov}.)
For  $W_\pm^+$, to compute the discrete part of the spectrum of the first term in (\ref{DVector1}) , we see that the $U(1)$ subgroup of $SL(2, {\mathbb R})$ has the charge $R_3 = b+1$, taking into account the intrinsic vector charge of $W_+^+$ 
and the curvature contribution. The corresponding  generic UIR of $SL(2, {\mathbb R})$ therefore has the extremal weight  $\Lambda = b +1 - k$ with $k \in {\mathbb Z}_+$. Putting this information together, we find
\beqa
- {\cal D}^{2 +}_{ V +} &=& \frac{1}{a^2}\left (-\Lambda (\Lambda -1) + R_3^2  - 2 b - 1  \right) \nn \\
&=& \frac{1}{a^2}\left ( - (b+1-k)(b+1 -k-1) + (b+1)^2 - 2 b-1 \right) \nn \\
&=& \frac{1}{a^2}\left ( -k(k-1) + 2 b k - b \right) \,, 
\label{DVector++D}
\eeqa 
where $k=0,1,2,...$ labels the Landau levels (LLs). The ground state, $|b+1,0 \rangle$, is specified by taking $k=0$ and has  negative energy $- \frac{b}{a^2}$. Representation theory of $SL(2, {\mathbb R})$ requires the condition $\Lambda = b+1-k \geq \frac{1}{2}$ to be fulfilled which gives $k \leq \lbrack b + \frac{1}{2} \rbrack$. This means that, for a given value of $b$, there are only as many LLs as allowed by this inequality and they are labeled by the integers $k$. In particular, contrary to the scalar case, there is a single discrete mode even at $b=0$. Clearly this is a zero mode. Following the same reasoning, the discrete part of the spectrum of $- {\cal D}^2_{V}$ on $W_-^+$ follows from taking $R_3 = b -1$ and $\Lambda = b -1 - k$. This gives
\beqa
- {\cal D}^{2 +}_{V -} &=& \frac{1}{a^2}\left (-\Lambda (\Lambda -1) + R_3^2  + 2 b - 1  \right) \nn \\
&=& \frac{1}{a^2}\left ( - (b-1-k)(b-1 -k-1) + (b-1)^2 + 2 b -1 \right) \nn \\
&=& \frac{1}{a^2}\left ( -k(k+3) + 2 b k + 3 b - 2 \right) \,, 
\label{DVector-+D}
\eeqa 
with the ground state $| b - 1,0 \rangle$ carrying the energy $(3b-2)/a^2$. We note that the requirement $\Lambda = b-1-k \geq \frac{1}{2}$ gives $k \leq \lbrack b - \frac{3}{2} \rbrack$. Thus, there are no discrete states for $W_-^+$ if $b < \frac{3}{2}$. 

For the continuous part of the spectrum of $- {\cal D}^2_{V}$ we have
\beq
\mbox{Spec}_C( - {\cal D}^2_{V}) =  \frac{\lambda^2 + \frac{1}{4} + (b \pm 1)^2}{a^2} \mp \frac{2b}{a^2} -\frac{1}{a^2} 
= {1\over a^2} \left( {\lambda^2 + \frac{1}{4} + b^2}\right) \,. 
\label{cont-spec}
\eeq
This result shows that, contributions to the spectrum from the spin-magnetic field coupling and the spin-curvature coupling and the vector charges $\pm 1$ for $H^2$ cancel each other and the continuous part of the spectrum for $W^+_\pm$ over $H^2$ is the same as that of a scalar \cite{{KKN1},{Comtet:1984mm},{Comtet:1986ki}}.
The term $1/(4 a^2)$ is obviously related to the
Breitenlohner-Freedman bound
\cite{BF}.

In the appendix of \cite{KKN1}, we have briefly stated the techniques used in the derivation of the  
exact expressions for the density of states for the discrete and the continuous spectrum of states of scalar and spinor particles and referred to the existing original and the recent literature \cite{Comtet:1984mm}, \cite{Comtet:1986ki} and \cite{Perelomov} for full details. In order to determine the corresponding density of states for the fields $W^+_\pm$, all we need to perform is the shift $b \rightarrow b \pm 1$ in the density of states of the scalar field, which leads to the expressions given in the Table\,\ref{cont-eigen} and Table\,\ref{disc-eigen}. In particular, we note that, 
the density of states for the continuous spectrum remains the same as that for the scalars, since 
$\cos (2 \pi b )$ is periodic under $b \rightarrow b \pm 1$.
Thus both the eigenvalues and the density of states is the same as what was obtained
for scalars.

\section{Semiclassical Estimate of the Density of States}

For the analysis of the pair production effect on curved spaces, a crucial ingredient in the calculation of the trace of the logarithm of the energy eigenvalues of the relevant differential operator for charged particles is the knowledge of the density of quantum states at a given energy. In this paper and its prequel \cite{KKN1}, which treated the scalar and spinor particles, we have focused on the spaces of the form $M \times {\mathbb R}^{1,1} $, where $M$ can be $S^2$ or $H^2$. 
General expressions for the
density of states on these spaces are known. While these are fairly 
simple to obtain in the case of $S^2$,
the corresponding calculations for 
the harmonics on $H^2$ (or more generally sections of 
$U(1)$-bundles over the hyperbolic plane), which correspond to
the principal continuous series representations of
$SL(2, {\mathbb R})$, are rather more involved
requiring more sophisticated group theoretical techniques.
Although we already know
the exact density of states on $S^2$ and $H^2$ for spin $0$ and $\frac{1}{2}$
from the existing literature, and inferred the result for spin $1$ particles on $S^2$ and $H^2$ from the properties of the corresponding isometry groups, it is still desirable to have a semiclassical estimate of the density of states on such spaces to further our understanding and physical intuition.   
This is the subject of this appendix.

The key procedure is the following. We 
obtain the classical trajectories for a point-particle moving on the space of interest.
These trajectories will be labeled by a number of parameters. The volume element
of the phase space (divided by $(2\pi )^n$) can be evaluated on the set of all such
these trajectories,
trading the momenta for the parameters labeling the trajectories. The result is then
 the semiclassical
measure of the number of trajectories needed in a path integral formula for evaluating the
trace of the evolution kernel for the operator which serves as the Hamiltonian for the
trajectories. The key result is that the Plancherel measure, semiclassically,
is simply the symplectic volume evaluated on the classical trajectories.
 
Perhaps not so surprisingly, Hamilton-Jacobi theory can be successfully exploited for this purpose and, in fact, we are able to obtain semiclassical estimates for the density of harmonics over $S^n$ and $H^n$. In order to see how this can be achieved and to handle both cases (and also the flat case 
${\mathbb R}^n$) on essentially equal footing we may use the Friedmann-Robertson-Walker type parametrization of the
metric for $M$ as
\be
ds ^2 = \frac{d r^2 }{1 -k r^2} + r^2 d \Omega_{n-1}
\ee
where $k =0, \frac{1}{a^2}, - \frac{1}{a^2}$ for ${\mathbb R}^n$, $S^n$ and $H^n$, respectively, and $a$ is the radius/length parameter of the latter. $d \Omega_{n-1}$ is the ``solid-angle" in $n-1$-dimensions and it can be expressed using the standard hyper-spherical coordinates. In particular, the inverse metric $g^{-1}_{\mu \nu} = g^{\mu \nu}$ has the diagonal form
\be
g^{ \mu \nu} \equiv diag \big( 1- k r^2, \frac{1}{r^2}, \frac{1}{r^2 \sin^2 \theta_2}, \frac{1}{r^2 sin^2 \theta_2 \sin^2 \theta_3 }, \cdots \,,
\frac{1}{r^2 \sin^2 \theta_2 \sin^2 \theta_3 \cdots sin^2 \theta_{n-1}} \big) \,.
\ee
where $\mu, \, \nu = 1, 2, \cdots, n$.

We introduce the Hamilton's principal function on ${\mathbb R}^{1,1} \times M $ by $S (t, z, r, \theta_2, \theta_3\,, \cdots \theta_{n})$, with $t,z$ denoting the time and the spatial direction on 
${\mathbb R}^{1,1}$. Although it is not necessary for the main purpose of this appendix, we also assume a uniform electric field, $E$, in the $z$-direction. After these preparatory steps, the Hamilton-Jacobi equation for a charged scalar of mass $m$ can be written as
\be
- \left (- \frac{\partial S}{\partial t} - E z \right)^2 + \left (\frac{\partial S}{\partial z}\right)^2 
+(1-kr^2) \left (\frac{\partial S}{\partial r} \right)^2 + g^{ij} \frac{\partial S}{\partial \theta_i}\frac{\partial S}{\partial \theta_j} + m^2 = 0 \,,
\label{hjeq1}
\ee
where $i,j=2,3,\cdots n$. Canonical momenta can be written as usual
\be
p_t \equiv  \varepsilon = - \frac{\partial S}{\partial t} \,, \quad p_z = \frac{\partial S}{\partial z} \,, \quad p_r = \frac{\partial S}{\partial r} \,, \quad p_i = \frac{\partial S}{\partial \theta_i} \,.
\ee
From (\ref{hjeq1}), we immediately observe that the Hamilton-Jacobi equation is cyclic in $t$ and the ``azimuthal"-angle $\theta_{n}$, which gives the energy $\varepsilon$ and the ``azimuthal" angular momentum $p_{n}$ as two of a number of conserved quantities (others will be determined shortly) and we may factor $ S \equiv S (t, z, r, \theta_2, \theta_3\,, \cdots \theta_{n})$ as
\be
S = W(z, r, \theta_2, \theta_3\,, \cdots \theta_{n-1}) + p_n \theta_n - \varepsilon t \,,
\label{hjf1}
\ee
where $ W(z, r, \theta_2, \theta_3\,, \cdots \theta_{n-1}) $ can be identified as Hamilton's characteristic function. We can further separate variables by writing
$ W(z, r, \theta_2, \theta_3\,, \cdots \theta_{n-1}) = W_n (z) + {\tilde W}(r, \theta_2, \theta_3\,, \cdots \theta_{n-1})$.
Using this in (\ref{hjeq1}) we get the equations
\beqa
&&(1-kr^2) \left (\frac{\partial {\tilde W}}{\partial r} \right)^2 +
\frac{1}{r^2} \left (\frac{\partial {\tilde W}}{\partial \theta_2} \right)^2 + \frac{1}{r^2 \sin^2 \theta_2} \left (\frac{\partial {\tilde W}}{\partial \theta_3} \right)^2+ \cdots  + \frac{p_n^2}{r^2 \sin^2 \theta_2 \cdots \sin^2 \theta_{n-1}} = \lambda_1^2 \nn \\
&&\left (\frac{d W_n}{d z} \right)^2 - (\varepsilon - E z)^2 + m^2 = - \lambda_1^2 \,, 
\label{hjeq2}
\eeqa
where $\lambda_1^2$ is introduced as a separation constant. First equation in (\ref{hjeq2}) gives rise to two equations
after multiplying the entire equation by $r^2$ and introducing another separation constant, $\lambda_2^2$, and this pattern can be continued until a set of $(n-1)$ decoupled ordinary differential equations is obtained. This is equivalent to 
the separation ansatz
${\tilde W} (r, \theta_2, \cdots, \theta_{n-1} ) = W_1(r) + W_2(\theta_2) +
\cdots + W_{n-1}(\theta_{n-1})$.
The full set of decoupled equations are thus
\beqa
r^2(1 -k r^2) \left ( \frac{d W_1}{dr} \right)^2 &=& \lambda_1^2 r^2 - \lambda_2^2 \,, \nn \\
\sin^2 \theta_i  \left ( \frac{d W_i}{d \theta_i} \right)^2 &=& \sin^2 \theta_i \lambda_i^2 - \lambda_{i+1}^2 \,, \hskip .1in i = 2, 3, \cdots, (n-1)
\eeqa
where $\lambda_n \equiv p_n$.
We solve these equations as
\be
\frac{d W_1}{d r} = \pm \sqrt{\frac{\lambda_1^2 r^2 - \lambda_2^2}{r^2 (1-k r^2)}} \,, \quad \frac{d W_i}{d \theta_i} = \pm \sqrt{\lambda_{i}^2 - \frac{\lambda_{i+1}^2}{\sin^2 \theta_i}}  \,.
\label{Wi}
\ee
The phase space volume element in terms of the dynamical variables $(r,\theta_i, \theta_n, p_r, p_i, p_n)$ is
\be
\frac{\omega \wedge \omega \cdots \wedge\omega}{(2 \pi)^n} \equiv \frac{1}{(2 \pi)^n} d p_r \, d r \prod_{i=2}^{n-1} d p_i d \theta_i d p_n d \theta_n  \,.
\ee
The separation constants $(\lambda_1, \lambda_i)$ can be used instead of the momentum variables $p_r, p_i$ to express the phase space volume element. The transformation between the dynamical variables $(r,\theta_i, \theta_n, p_r, p_i, p_n)$ and $(r,\theta_i, \theta_n, p_r, \lambda_i, p_n)$ 
leads to the Jacobian matrix $J$, whose determinant is not identity, and the phase space volume element takes the form
\beqa       
\frac{\omega \wedge \omega \cdots \wedge \omega}{(2 \pi)^n} &\equiv& \frac{1}{(2 \pi)^n} |\det J| \, d p_r dr \prod_{i=2}^{n-1} d \lambda_i d \theta_i d p_n d \theta_n \,, \nn \\
&=& \frac{1}{(2 \pi)^n} \frac{\lambda_1 d \lambda_1}{\sqrt{\lambda_1^2 r^2 - \lambda_2^2}} \frac{r dr }{\sqrt{1-k r^2}} \prod_{i=2}^{n-1} \frac{\lambda_i d \lambda_i d \theta_i}{\sqrt{\lambda_i^2- \frac{\lambda_{i+1}^2}{\sin^2 \theta_i}}} d p_n d \theta_n \,.
\eeqa
Performing the integrals over $\lambda_i$ and $p_n$ in the reverse order, that is starting with the integral over $p_n$, and following this up with integrals over $\lambda_i$ with $i$ from $n-1$ to $2$, we have
\beqa
\int_{-c}^c
\frac{d p_n}{\sqrt{\lambda_{n-1}^2 - \frac{p_n^2}{\sin^2 \theta_{n-1}}}} &=& \pi |\sin \theta_{n-1}| \,, 
\hskip .2in c = {\sqrt{ \lambda^2_{n-1} \sin^2\theta_{n-1}}}\, ,\nn \\ 
\int |\lambda_i|^{n-i-1} \frac{ \lambda_i d \lambda_i}{\sqrt{\lambda_{i-1}^2- \frac{\lambda_{i}^2}{\sin^2 \theta_{i-1}}}} &=& |\sin \theta_{i-1}|^{n-i+1} |\lambda_{i-1}|^{n-i} \frac{1}{2} \frac{\Gamma(\frac{n-i+1}{2}) \Gamma(\frac{1}{2})} {\Gamma(\frac{n-i+2}{2})} \,, \quad i = 3, \cdots n-1 \,, \nn \\
\int |\lambda_2|^{n-3} \frac{\lambda_2 d \lambda_2}{\sqrt{\lambda_1^2 r^2 - \lambda_2^2}} &=& |\lambda_1|^{n-2} r^{n-2} \frac{1}{2} \frac{\Gamma(\frac{n-1}{2}) \Gamma(\frac{1}{2})}{\Gamma(\frac{n}{2})} \,,
\eeqa
where the integrations over the $\lambda_i$, $i= 2, \cdots, (n-1)$,  are from zero to the positive
turning points of the integrands. The phase space element can be expressed as 
\beqa
\frac{\omega \wedge \omega \cdots \wedge\omega}{(2 \pi)^n} &\equiv & \frac{1}{(2 \pi)^n}\, \lambda^{n-1} \, d \lambda \, \pi \, 2^{n-1} \, (\frac{1}{2})^{n-2} \, \prod_{i=2}^{n-1} \frac{\Gamma(\frac{n-i+1}{2}) \Gamma(\frac{1}{2})} {\Gamma(\frac{n-i+2}{2})}\, d V \,, \nn \\
&=& \lambda^{n-1} d \lambda \,{1\over (2 \pi)^n} {2\, \pi^{n\over 2}\over \Gamma({n \over 2})}\, dV\,, \nn \\
&=:& {\mathbb P}_n \, (\lambda) \,d V \,.
\label{pn}
\eeqa
We have included an additional factor
$ 2^{n-1}$ in this expression. This can be viewed as accounting for both the
signs in the expressions for $W_i$, $i = 1, 2, \cdots, (n-1)$ as in (\ref{Wi}).
Equivalently, we may think of this as being due to the 
 two-fold orientation of the paths in each case.
We have also dropped the subscript in $\lambda_1$, since after the integrations over $\lambda_i$ ($i=2,\cdots, n-1$), this is the only variable remaining in the phase space measure apart from the configuration space volume element $d V$. The latter is    
\beqar
d V &=& \frac{r^{n-1}}{\sqrt{1- k r^2}} \,\vert\sin \theta_2^{n-2} \sin \theta_3^{n-3} \cdots \sin^2 \theta_{n-2} \sin{\theta_{n-1}} \vert\, dr d \theta_2 d \theta_3 \cdots d \theta_{n}\nonumber\\
&=& \frac{r^{n-1}}{\sqrt{1- k r^2}}\, d r \, d \Omega_{n-1}\,.
\eeqar

The quantity ${\mathbb P}_n(\lambda)$ introduced in the last line of (\ref{pn}) can be identified as our semiclassical estimate for the density of harmonic functions on $S^n$ and $H^n$ in terms of a continuous parameter, $\lambda$, which can naturally be thought as a wave-number. For comparison with group theoretical results, it is useful to express ${\mathbb P}_n$ for even and odd values of $n$ separately. We have
\be
{\mathbb P}_{2m}(\lambda) = \frac{\lambda^{2m-1} d \lambda}{(2 \pi)^m (2m-2)!!} \,, \quad {\mathbb P}_{2m+1}(\lambda) = \frac{\lambda^{2m} d \lambda}{(2 \pi)^m \,\pi \, (2m-1)!!} \,.
\ee

For $S^n = \frac{SO(n+1)}{SO(n)}$, the density of harmonic functions can be given as a sum over the dimensions of the irreducible representation $(l,0\,,\cdots\,,0)$ (in the Dynkin notation) of $SO(n+1)$ divided by the volume of $S^n$, i.e.
\beqa
{\mathbb P}_{S^n}  &\equiv& \sum_{l \geq 0} \frac{\mbox{dim}(l,0\,,\cdots\,,0)}{\mbox{vol}(S^n)} = 
\sum_{l \geq 0} \frac{(l+n-2)! \, (2l +n-1)}{2 \pi^{(n+1)/2} \, l! \,(n-1)! \,  a^n} \Gamma (\frac{n+1}{2}) \nn \\
&\underset{(l/a) \rightarrow \lambda}{\longrightarrow} & \frac{\lambda^{n-1} \, d \lambda}{\pi^{(n+1)/2} \, (n-1)! } \Gamma (\frac{n+1}{2}) 
= 
\begin{cases}
{\mathbb P}_{2m}(\lambda) & n = 2m \,, \\
{\mathbb P}_{2m+1}(\lambda) & n= 2m+1 \,. 
\end{cases}
\eeqa
From the second line of this expression, we infer that the semiclassical formulae ${\mathbb P}_{2m}(\lambda)$ and ${\mathbb P}_{2m+1}(\lambda)$ obtained from the Hamilton-Jacobi theory give estimates, which are in complete agreement with the $(l/a) \rightarrow \lambda$ limit of the exact result. 

For the hyperbolic spaces $H^n$ the situation is a little more intricate. In this case, the exact result for the density of harmonic states is given in terms of the Plancherel measure, which can be obtained after rather tedious considerations of the properties of the non-compact group $SO(n,1)$ as \cite{Perelomov}
\be
{\mathbb M}_{H^{2m}} = \frac{|\Gamma(m-\frac{1}{2} +i \lambda)|^2}{|\Gamma(i\lambda)|^2} \,, \quad {\mathbb M}_{H^{2m+1}} = \frac{|\Gamma(m + i \lambda)|^2}{|\Gamma(i\lambda)|^2} \,, 
\ee
Multiplying these with the phase space factor $\frac{1}{(2 \pi)^n}$, the total solid angle $\frac{2 \pi^{n/2}}{\Gamma(\frac{n}{2})}$, the differential element $d \lambda$ and simplifying the expressions involving the $\Gamma$-functions, we have
\beqa
{\mathbb P}_{H^{2m}}(\lambda) &=& \frac{1}{(2 \pi)^{2m}} \frac{2 \pi^m}{\Gamma(m)} \lambda \tanh \pi \lambda ~  d  \lambda   \prod_{k=2}^m \left[(k-\frac{3}{2})^2 + \lambda^2 \right] , \quad
m \geq 2
  \nn \\
&\underset{\lambda \rightarrow \infty}{\longrightarrow}& \frac{\lambda^{2m-1} d \lambda}{ (2\pi)^m (2m-2)!!} = {\mathbb P}_{2m}(\lambda) \, \\
{\mathbb P}_{H^{2}}(\lambda)&=&{1\over 2 \pi} \lambda \tanh\pi\lambda~d\lambda 
~~\underset{\lambda \rightarrow \infty}{\longrightarrow}~~
{\lambda d \lambda \over 2 \pi} = {\mathbb P}_{2}(\lambda)
\eeqa
and
\beqa
{\mathbb P}_{H^{2m+1}}(\lambda) &=& \frac{1}{(2 \pi)^{2m+1}} \frac{2 \pi^m \sqrt{\pi}}{\Gamma(m+\frac{1}{2})} d  \lambda \prod_{k=1}^m \left [(k-1)^2 + \lambda^2 \right]  \nn \\
&\underset{\lambda \rightarrow \infty}{\longrightarrow}& \frac{\lambda^{2m} d \lambda}{(2 \pi)^{m} \pi (2m-1)!!} = {\mathbb P}_{2m+1}(\lambda) \,.
\eeqa
Thus, we see that the semiclassical estimates obtained from the Hamilton-Jacobi theory matches with the large $\lambda$ limit of the exact results on $H^{2m}$ and $H^{2m+1}$. 
For odd dimensions, the difference between the exact and semiclassical
results is a polynomial of lower order in $\lambda$, while for even dimensions
there are similar polynomial corrections and
also corrections of the type $\lambda^{2 m-1} e^{- 2 \pi \lambda}$ (and further subdominant terms), due to the
$\tanh \pi \lambda$ factor.

We have not discussed the second equation in (\ref{hjeq2}) which deals with the $z$-dependence
of the action.
This can be used to calculate a tunneling amplitude as in the usual WKB analyses, giving
a semiclassical estimate of the pair production rate.
We have not done this, since, in the main text,
we have already calculated the rate without making such an approximation, focusing instead on the
Plancherel measure.

\vskip 1em

\noindent{\bf \large Acknowledgments}

\vskip 1em

\noindent S.K.'s work was carried out during his sabbatical stay at the physics department of CCNY of CUNY and he thanks V.P. Nair and D. Karabali for the warm hospitality at CCNY and the metropolitan area. S.K. also acknowledges the financial support of the Turkish Fulbright Commission under the visiting scholar program.  
The work of VPN was supported in part by the U.S.\ National Science
Foundation grant PHY-1820721. DK and VPN acknowledge the support of PSC-CUNY awards.

\vskip 1em

\end{document}